# Spectral Indices in Cooling Flow Galaxies: Evidence for Star Formation

N. Cardiel,[1] J. Gorgas,[1] and A. Aragón-Salamanca[2]
[1] *Departamento de Astrofísica, Facultad de Ciencias Físicas, Universidad Complutense de Madrid, 28040 Madrid, Spain*
[2] *Institute of Astronomy, Madingley Road, Cambridge CB3 0HA, United Kingdom*

**ABSTRACT**
Spectroscopic observations of central dominant cluster galaxies, with and without cooling flows, are presented. Through the analysis of absorption spectral features, namely the strength of the magnesium absorption at 5175 Å and the 4000 Å break, both in the galaxy centres and as a function of radius, we have been able to estimate the ongoing star formation induced by the large amounts of gas accreted onto cooling-flow galaxies. A correlation between the central spectral indices and the mass accretion rate is found in the sense that galaxies located in clusters with large cooling flows exhibit lower $Mg_2$ and $D_{4000}$ indices. A similar correlation with $D_{4000}$ was previously reported by Johnstone, Fabian & Nulsen (1987). Our work, with the inclusion of the correlation in $Mg_2$, adds further weight to the conclusion that these spectral anomalies are caused by recent star formation. The application of simple stellar population models reveals that the measured indices are explained if a relatively small fraction of the total mass flow (5–17%) is forming new stars with a normal initial mass function. However, we argue that this is only a lower limit, and conclude that a large fraction of the gas accreted *inside* the galaxy could be forming stars. We find that spectral gradients in some cooling flow galaxies flatten in the internal regions ($r \gtrsim r_e$), where emission lines are usually detected. Gradients measured in the inner galaxy regions are, in the mean, lower than those of normal ellipticals, and exhibit a hint of a correlation with $\dot{M}$. Application of the same population models to the observed spectral gradients allows us to conclude that the ongoing star formation is concentrated towards the inner parts of the cooling flow galaxies and, therefore, the star formation does not follow the X-ray derived mass accretion profiles. Simultaneously, the spectral indices in the outer regions of some galaxies with *and* without cooling flow attain extremely low values, suggesting that they could be hosting star formation with an origin not related to the cooling flows.

## 1 INTRODUCTION

### 1.1 Cooling flows

X-ray observations indicate that a substantial fraction of the giant elliptical galaxies found at the centres of galaxy clusters are accreting large amounts of gas (cooling flows) from the intracluster medium at typical rates of $\dot{M} \sim 100 M_{\odot}$/yr (Fabian, Nulsen & Canizares 1984). The study of an X-ray flux-limited sample of galaxy clusters (Edge, Stewart & Fabian 1992) has shown that central cooling times are less than the Hubble time in at least 70% and, possibly, 90% of the clusters, suggesting that cooling flows should be long-lived. If this is the case, the brightest cluster galaxies found at the centres of such flows could have accreted total masses of the order of $10^{11}$–$10^{12} M_{\odot}$, comparable with their estimated luminous masses. X-ray data with good spatial resolution indicates that the mass deposition profiles are reasonably fitted by power-laws of the form $\dot{M}(< r) \propto r^n$, where $n$ (the drop-out parameter) is typically $n \simeq 1$, implying most of the gas is not flowing to the very centre of the cluster (Thomas, Fabian & Nulsen 1987). For extensive reviews on cooling flows, see Sarazin (1986), Fabian, Nulsen & Canizares (1991), and Fabian (1994).

Although alternative scenarios have also been proposed to reduce the large amounts of cooled gas flowing to the cluster centres, they can not successfully explain the X-ray data (Fabian et al. 1991 and references therein). A challenging possibility is the existence of *heating flows* (de Jong et al. 1990; Sparks 1992). However, Fabian, Canizares & Böhringer (1994) have shown that the *heating flow* scenario is inconsistent with X-ray imaging observations and requires implausible stability and boundary conditions.

The fact that cooling flows could be common and long-lived processes in X-ray clusters of galaxies strongly suggests that the origin and evolution of such flows should be related to the history and evolution of the clusters. However, analysing the X-ray emission from bimodal clusters, Henriksen (1993) has argued that cooling flows do not exist over the whole lifetime of the clusters, since merger processes should disrupt the flows.

One of the outstanding questions in the understanding of the cooling flow phenomenon is what is the final fate of the accreted gas. Despite the fact that it cannot be completely ruled out that the accreted gas could stay in an almost undetectable gaseous form, it is generally thought that the more plausible fate for most of the gas is the formation of new stars (eg. O'Connell & McNamara 1989; Fabian et al. 1991). However, recent observations have revealed that many cooling flow galaxies may contain cold gas at some extent (McNamara, Bregman & O'Connell 1990; White et al. 1991). This result implies that star formation in cooling flows may not be as efficient as previously thought.

### 1.2 Previous work

If star formation is taking place at the rates inferred from X-ray observations, and with a 'normal' (i.e. solar neighbourhood) initial mass function (IMF), the accreting galaxies would be remarkably blue, even if $\dot{M}$ were substantially overestimated. The first studies devoted to detect this star formation did not find significant colour anomalies in cooling flow galaxies (Lachièze-Rey, Vigroux & Souviron 1985; Romanishin 1986), except for NGC 1275 and PKS 0745−191 (Romanishin 1987), two galaxies with prominent accretion rates. Later photometric work in the near-infrared and optical ranges (Thuan & Puschell 1989; McNamara & O'Connell 1992) have confirmed that galaxies with large flows exhibit bluer central colours than non-accreting cD galaxies, finding a possible correlation between colour and $\dot{M}$. Furthermore, colour anomalies are found to be more spatially extended as the accretion rate increases. In general, the blue anomalies imply that only a few percentage ($< 10\%$) of the accreted gas may be forming stars with a normal IMF. Recently, McNamara & O'Connell (1993) have found that the blue colours observed in A 1795 and A 2597 are located in patches coincident with the radio lobes of the galaxies, suggesting a connection between the radio sources and the star formation (or, at least, the source of the blue light).

Spectroscopic work on cooling flow galaxies has led to similar conclusions as the photometric studies. Although Hu, Cowie & Wang (1985) did not find evidence for star formation, work from Wirth, Kenyon & Hunter (1983), Bertola et al. (1986) and Crawford et al. (1989), employing population synthesis methods, have shown that, in order to reproduce the continuum spectral distribution in the optical and ultraviolet, star formation with a normal IMF should be taking place at rates about 5% of the total accretion flow. An important result is that presented by Johnstone, Fabian & Nulsen (1987, hereafter JFN), who reported a correlation of the $\lambda$4000 Å break ($D_{4000}$; see definition in Bruzual 1983) with the mass accretion rate, in the sense that $D_{4000}$ decreases for large accretion flows. Furthermore, JFN, McNamara & O'Connell (1989), and Allen et al. (1992) found a clear correlation between the blue excess and the emission line luminosities (A discussion of emission lines in cooling flow galaxies is presented in Section 5). We must highlight the recent work of Crawford & Fabian (1993), in which it is noted that the blue spectral anomalies can also be explained by power-law spectra, thus suggesting that the blue colours could be caused by a non-thermal source instead of star formation. These authors also remark that spectral distributions in cooling flow galaxies are more consistent with a past starburst than with a constant star formation from the cooling flow.

Summarising, the main conclusion from the former photometric and spectroscopic works is that only a small percentage ($\leq 10$ %) of $\dot{M}$ may be forming stars with a normal IMF. The fate of the remaining gas remains unclear. Several authors (Sarazin & O'Connell 1983; O'Connell & McNamara 1989; Fabian et al. 1991) have argued that the most viable repository is low-mass stars, or even planetoids. Actually, an IMF biased to low masses could be favoured by the the lack of dust in the cooled gas and the absence of large molecular clouds. Ferland, Fabian & Johnstone (1994) have shown that the Jeans mass in a cooling flow (mainly at large radii) can be $\leq 0.1$ $M_\odot$, thus favouring the formation of low-mass stars. Recently, Schombert, Barsony & Hanlon (1993), from optical and near-IR imaging, find some evidence for that low-mass star population. In addition, Johnstone & Fabian (1989) pointed out that the observation of a giant red envelope galaxy (GREG; Maccagni et al. 1988) can be interpreted as a population of low-mass stars formed from a cooling flow.

Although the nuclear colour anomalies of galaxies with prominent cooling flows have been firmly established, the observed effects of gas flows in the colour gradients of the accreting galaxies are quite unclear. Apart from NGC 1275 and PKS 0745−191, which exhibit large positive colour gradients (Romanishin 1987) consistent with the star formation concentrated towards the nuclear regions, the studies of colour gradients in cooling flow galaxies have led to contradictory results. Whilst some authors do not find significant differences in gradients compared to giant elliptical galaxies (Hu et al. 1985; Lachièze-Rey et al. 1985; Peletier 1992), other work seems to indicate that colour gradients may flatten as mass accretion rate increases (Mackie, Visvanathan & Carter 1990; McNamara & O'Connell 1992), though the scatter of that correlation is too large to confirm it. The red upturns found by McNamara & O'Connell (1992) in the outer parts of some brightest cluster galaxies, both with and without cooling flows, are not consistent with other studies (eg. Peletier 1992).

### 1.3 Spectral indices in cooling flow galaxies

The study of spectral absorption features in cooling flow galaxies can shed new light on the understanding of the possible star formation in these kind of objects (Sarazin 1986). This approach was first followed by Gorgas, Efstathiou & Aragón-Salamanca (1990, hereafter GEA). They measured $Mg_2$ indices in the central regions of a sample of brightest cluster galaxies which includes some cooling flow objects. Since line-strength gradients are usually found in early-type galaxies, the conclusions of that work depended on the unknown size of the gradients for the cooling flow objects (needed to perform aperture corrections). Under the assumption that these gradients were similar to those found in ellipticals, we concluded that no sign of star formation were apparent in the optical spectra of cooling flow galaxies. This result, in contradiction with the findings of JFN, can not be confirmed until a systematic study with spatial resolution of spectral indices in brightest cluster galaxies is performed. This is one of the aims of this paper. Simultaneously, the determination of spectral gradients is also very important in order to get information about the radial distribution of the possible ongoing star formation.

In this paper we focus on the study of the $D_{4000}$ and $Mg_2$ spectral indices in the centre and along the radii of a sample of brightest cluster galaxies with and without cooling flows. These indices have proved to be powerful indicators of changes in the stellar populations of early-type galaxies. The break at $\lambda$4000 Å has been widely used as an indicator of spectral evolution in high-redshift galaxies (eg. Hamilton 1985; Dressler 1988; Charlot & Silk 1994). The index is quite sensitive to the temperature of the main sequence turnoff point, although its dependence with metallicity, reported as negligible by Dressler & Shectman (1987), could be significant (Worthey 1994). The behaviour of the $Mg_2$ index (see Faber et al. 1985 for a definition) with stellar atmospheric parameters has been extensively studied by Gorgas et al. (1993). Obviously, it is also sensitive to metallicity. In order to avoid this problem, throughout this work we will assume that metallicity variations along cooling flow galaxies are similar to those for ellipticals. Both indices are suitable to be measured at relatively low signal-to-noise ratios, allowing the determination of spectral indices far from the galaxy centres.

The galaxy sample, data reduction techniques and error estimation are described in Section 2. The spectral indices in the central regions of the galaxies are presented in Section 3, together with the implications on the star formation from cooling flows. In Section 4, the results and interpretation of spectral gradients are given. Section 5 is devoted to a discussion of the emission lines found in some of the galaxies in the sample, and their connection to the results of previous sections. Finally, our conclusions are summarised in Section 6.

## 2 OBSERVATIONS AND DATA REDUCTION

### 2.1 The observations

Long-slit spectroscopic observations of brightest cluster galaxies were carried out in three runs with two different telescopes. In the first two runs (February and November, 1991) we observed with the 4.2$m$ WHT at the Roque de los Muchachos Observatory (La Palma, Spain) employing the ISIS double spectrograph mounted at the $f/11$ Cassegrain focus. Both arms of ISIS were used in the first run, using the IPCS II detector in the blue arm and a CCD in the red. Due to linearity problems in the IPCS II (see below) we changed this detector by a blue coated CCD for the second run. Here we also observed with the red arm, though weather conditions prevented these exposures from being useful, except for confirming the existence of some emission lines.

The observations of the third run (June 1993) were carried out at the 3.5$m$ telescope at the German-Spanish Astronomical Observatory at Calar Alto (Almería, Spain). We used the Cassegrain Twin Spectrograph with a blue coated TEK CCD in the red channel. The observations took place in grey nights.

Details of the observational configurations for each run are given in Table 1. In general, we obtained intermediate-resolution spectra from the 4000 Å break to the H$\alpha$ line.

We observed a sample of 10 central dominant galaxies, comprising 8 cooling flow galaxies and 2 in which no cooling flow has been detected. These galaxies were chosen to spread a range in mass deposition rates and redshifts. It must be noted that one of the non-cooling flow galaxies (NGC 7728) is a strong-radio source (see, e.g., Ball, Burns & Loken 1993). The sample is listed in Table 2 together with additional information including total exposure times and position angles of the spectrograph slit. The effective exposure times for PKS 0745−191 and A 644 were not as large as those required to obtain good signal-to-noise spectra outside the galaxy central regions. We also took spectra of a sample of G–K giant stars to be used as templates for velocity dispersion measurements. These stars will also be used to confirm that our $Mg_2$ indices are in the correct photometric system, as explained in the next section.

### 2.2 Data Reduction and Error Analysis

The data reduction process is based in that described by Gorgas (1987) and GEA. Initial chip's cosmetic (essentially bias, dark current subtraction and cosmic ray cleaning) was performed with the standard routines of the FIGARO package. Flatfielding, wavelength calibration, C-distortion correction, sky subtraction, extinction corrections, centering and binning of the spectra, and spectral indices measurements were achieved with the aid of own reduction programs.

In order to get reliable measurements of the indices, especially at low surface brightness levels, it is necessary to follow in detail the error propagation throughout the data reduction. The use of our reduction package allowed us to control errors more carefully than with standard software. We present a summary of some of the reduction steps, the sources of error and their main effects.

Lamp and dome flat-fields were employed to correct pixel-to-pixel differential response, whereas twilight flat-fields were used to compensate for two-dimensional low-frequency scale sensitivity variations on the chip. Unfortunately, flat-fields of Run 3 revealed that an electronic interference noise pattern was present in all the frames up to a level of 1% in the number of counts/pixel. The measured pattern exhibited unpredictable spatial shifts along the night that prevented us from removing it. Nevertheless we estimated its effect in the index measurements by introducing additional noise (with similar frequency ranges and random spatial shifts) in the final spectra. The derived errors, which turned out to be $\simeq 0.004$ mag for $Mg_2$ and $\simeq 0.02$ for $D_{4000}$, were taken into account in the calculation of the total errors.

In order to check the linearity of the detectors, we carried out a comprehensive analysis of several flat and arc frames taken, with that purpose, with different slit widths and exposure times. As it was expected, the IPCS II detec-tor showed important linearity deviations. For instance, at a count rate level of 0.2 Hz (counts/pixel/sec) about 9% of counts are lost. We found that this behaviour depends on the image structure, being thus quite difficult to correct. Anyway, the count rates in the galaxies observed with this detector are low enough to assume that the effect is not affecting the index measurements, as can be checked comparing the results for the galaxy (A 1126) observed with both the IPCS II and the CCD ($\mathrm{Mg}_2^{r_e/20}$(IPCS II) $= 0.291 \pm 0.015$, $\mathrm{Mg}_2^{r_e/20}$(CCD) $= 0.283 \pm 0.007$). On the other hand, the tests showed that the CCD's used in Runs 2 and 3 showed negligible non-linearity effects up to 60000 counts. The CCD employed in Run 3 exhibited small departures from linearity at high number of counts (e.g. at $N_c$=30000 overcounts $\sim$ 2%). In this case, the observations were corrected from this effect, although it marginally affects only the giant star spectra.

**Table 1.** Observational configuration.

| | Run 1 | | Run 2 | Run 3 |
|---|---|---|---|---|
| Dates | 16 Feb 1991 | | 1–3 Nov 1991 | 8–10 Jun 1993 |
| Telescope | WHT 4.2m | | WHT 4.2m | CAHA 3.5m |
| Spectrograph | ISIS blue | ISIS red | ISIS blue | CTS |
| Detector | IPCS II | CCD EEV2 | CCD EEV6 | CCD TEK 12 |
| Dispersion | 1.21 Å/pixel | 2.72 Å/pixel | 2.70 Å/pixel | 3.46 Å/pixel |
| Wavelength Range | 3450–6650 Å | 4600–7700 Å | 3500–6850 Å | 3750–7350 Å |
| Spectral Resolution | 17 Å | | 17 Å | 8 Å |
| Slit Width | 2.05 arcsec | | 2.0 arcsec | 2.1 arcsec |
| Spatial Scale | 1.20 arcsec/pixel | 0.34 arcsec/pixel | 0.34 arcsec/pixel | 1.79 arcsec/pixel |

**Table 2.** Galaxy sample.

| Cluster | Galaxy | R.A. (1950) | Decl. (1950) | z | Run | Exposure (secs) | P.A. | Comments |
|---|---|---|---|---|---|---|---|---|
| A 262 | NGC 708 | 01 49 50.0 | +35 54 22 | 0.0163 | 2 | 9000 | 40° | |
| 2A 0335+096 | ... | 03 35 57.0 | +09 48 28 | 0.0346 | 2 | 10800 | 144° | clouds |
| PKS 0745−191 | ... | 07 45 18.0 | −19 10 14 | 0.1028 | 1 | 2000 | 55° | blue arm |
| A 644 | ... | 08 14 59.0 | −07 21 23 | 0.0704 | 2 | 6355 | 16° | clouds |
| A 1126 | ... | 10 51 11.0 | +17 07 01 | 0.0828 | 1 | 6750 | 120° | clouds, blue arm |
| | | | | | 1 | 3000 | 120° | red arm |
| A 1795 | ... | 13 46 34.2 | +26 50 25 | 0.0636 | 3 | 12570 | 15° | |
| A 2124 | ... | 15 43 05.7 | +36 15 54 | 0.0671 | 3 | 12200 | 160° | moon |
| A 2199 | NGC 6166 | 16 26 55.7 | +39 39 39 | 0.0314 | 3 | 9600 | 30° | moon |
| A 2319 | ... | 19 19 36.7 | +43 51 00 | 0.0529 | 3 | 12521 | 0° | moon |
| A 2634 | NGC 7728 | 23 36 00.0 | +26 45 01 | 0.0315 | 2 | 3285 | 10° | |

Prior to the wavelength calibration, arc frames were employed to correct from C-distortion in the images. This rotation correction guaranteed alignment errors below 0.1 pixels. Spectra were converted to a linear wavelength scale using typically 50 arc lines fitted by $5-7^{\rm th}$ order polynomials, with rms errors of $0.5-1$ Å. These translate to negligible uncertainties in the indices ($\Delta\mathrm{Mg}_2 \simeq 0.0015$, $\Delta D_{4000} \simeq 0.001$). In addition, spectra taken with the IPCS II were corrected from S-distortion.

Since in the outer parts of some galaxies we are measuring indices in spectra with light levels of only a few percent of the sky signal, the sky subtraction is a critical step in our data reduction. For each galaxy observation a sky image was generated fitting for each channel (pixel in the $\lambda$ direction) a low order polynomial to regions selected at both sides of the galaxy. It is impossible to obtain, *a priori*, a good determination of the quality of the sky subtraction. However, an upper limit for this error can be evaluated by subtracting sky images with light levels altered by several factors. When the introduced factors are excessively large, sky lines, badly removed, appear in the galaxy spectra. In this way, we estimate that sky subtraction errors are, in most cases, well below 1%. In order to visualise the effect of a wrong sky subtraction, in Figure 1 we have plotted the $D_{4000}$ index along the radius for the galaxy in A 2199 varying the sky level between $\pm 1$%, using a constant step of 0.1%. As it could be expected, the measured indices in the outer regions increase dramatically when the sky is overestimated whereas the indices are moderately lower when the sky level is reduced. This figure shows that the low indices attained at the outer parts of some galaxies of the sample are very unlikely due to an underestimation of the sky, since this would imply high systematic errors that should be clearly detected as unremoved sky lines. A possible systematic overestimation of the sky level could arise if the galaxy contribution to the regions from where the sky is extracted is not negligible. To explore this effect we have used surface brightness profiles from the literature (Schombert 1986, 1987; Romanishin & Hintzen 1988) to estimate the relative contribution of the galaxy. When necessary, the effect has been taken into account subtracting from the sky spectra a scaled and averaged galaxy spectrum, though the effect on the measured indices is quite small (compared with the size of the rest of the errors) even in the outer spectra.

It is interesting to note in Figure 1 that random errors in the sky subtraction produce an asymmetric distribution of the $D_{4000}$ errors. We have investigated this effect in more detail by running numerical simulations of the sky subtraction in synthetic indices. Figure 2 plots a histogram of the $D_{4000}$ measured in simulations with a theoretical value of 2.4 (solid line) for a galaxy spectrum at a 10% level of the sky brightness, when we introduce a 1% Gaussian error in the sky level. The mean index (dashed line) is above the true value, whereas the mode of the distribution is below the actual index. This skewed distribution has been taken into account to estimate the errors due to photon noise and, in general, should be considered when measuring indices in galaxy spectra close to the sky brightness (e.g. high-redshift galaxies). Similar results were obtained in simulations with the $Mg_2$ index.

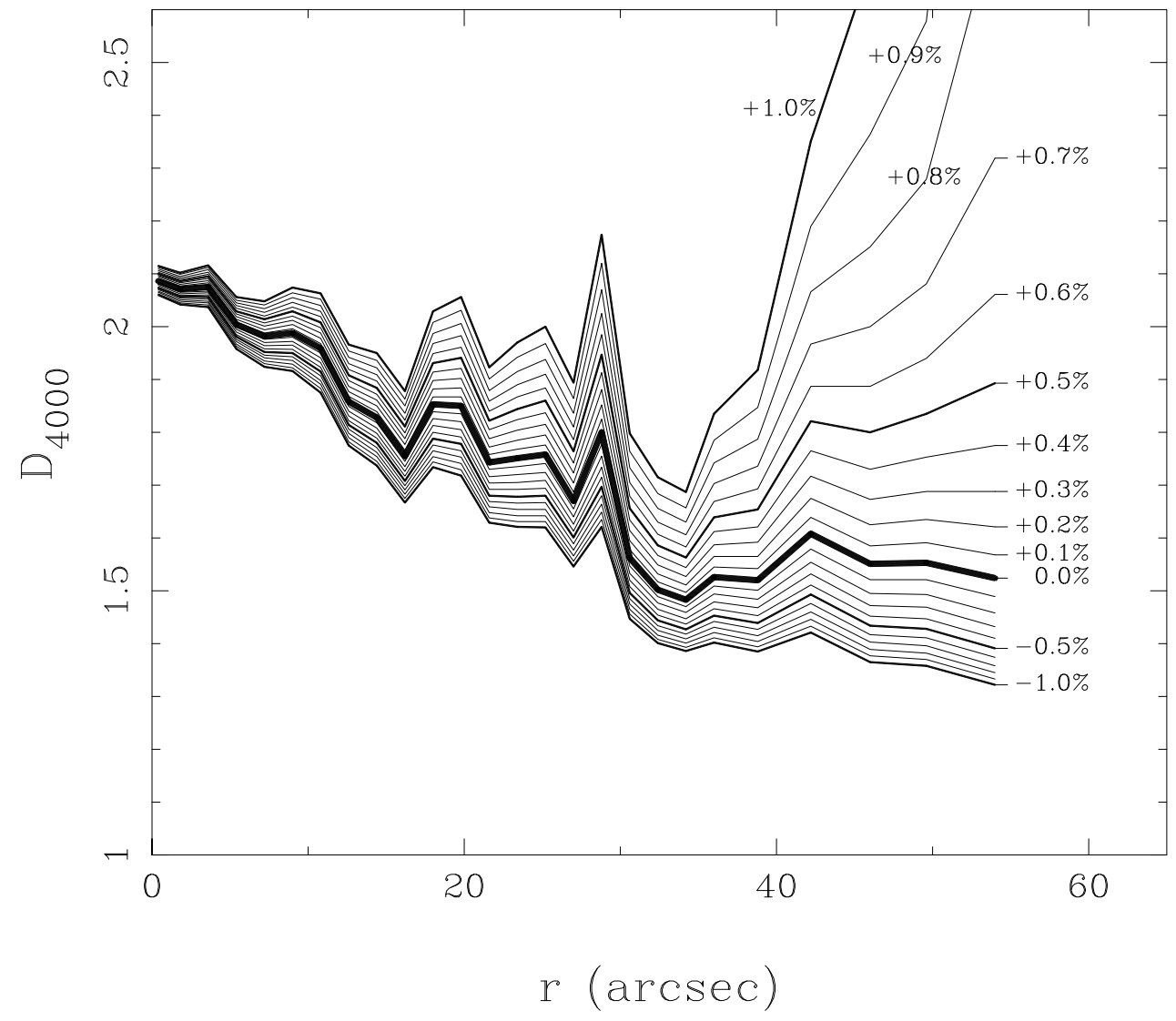

**Figure 1.** Effects of the sky subtraction errors in the $D_{4000}$ gradient of A 2199. The thick line is the measured gradient, whereas thin lines represent the gradient we would measure if the sky level is changed between $\pm1\%$ with a constant step of 0.1%. When the sky level is overestimated, indices in the outer galaxy regions grow dramatically, whereas underestimating the sky level leads to smaller changes in the index.

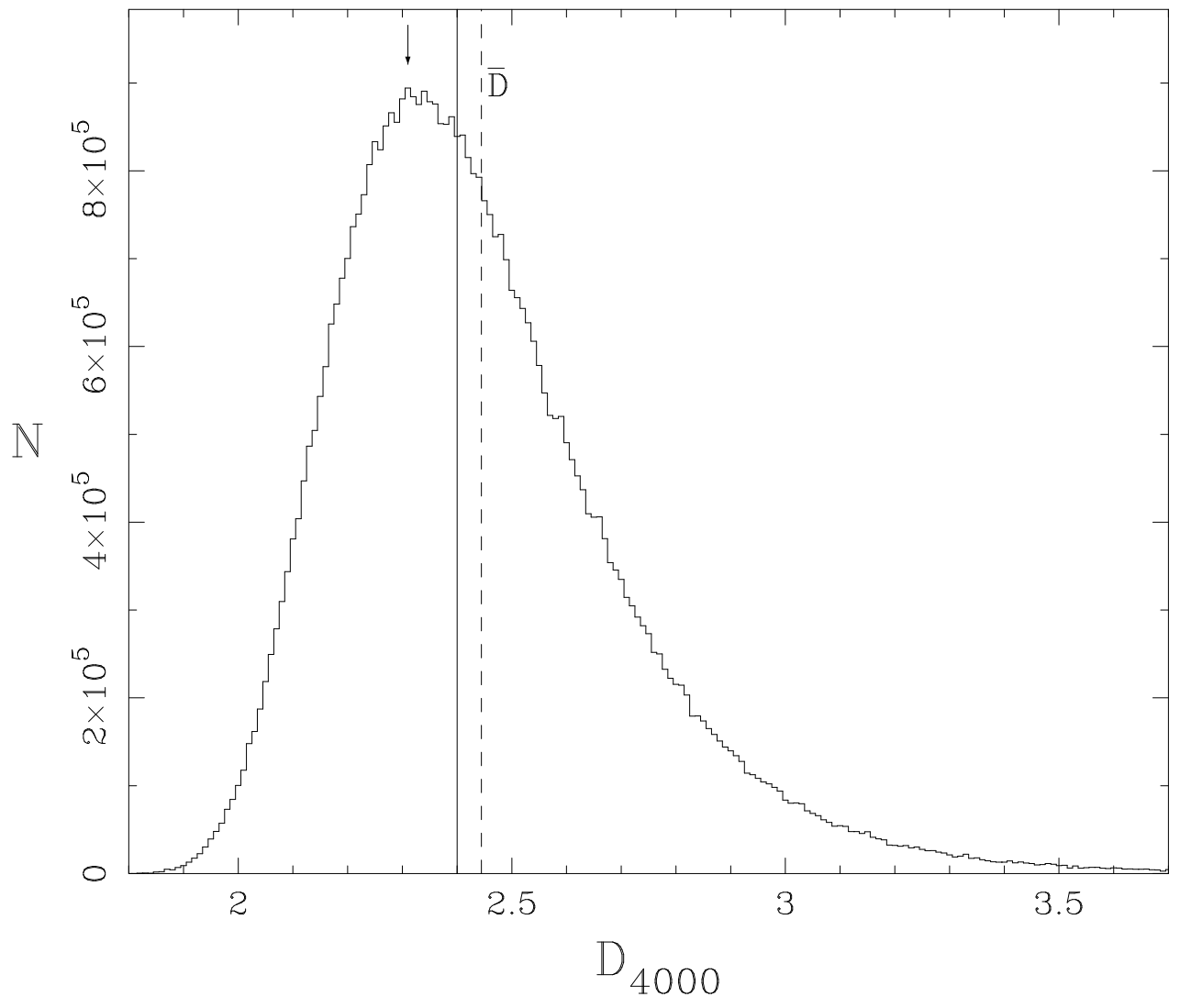

**Figure 2.** Histogram of the $D_{4000}$ values obtained in numerical simulations of the spectral index measured in a galaxy spectrum that has only 10% of the sky brightness, when introducing a 1% Gaussian error in the sky level and considering a theoretical value $D_{4000}=2.4$ (solid line). The mean index (dashed line) is above the true value, whereas the mode of the distribution (arrow) is below the real index.

Atmospheric extinction was calculated using the extinction curve of King (1985) for Run 1 and 2, and the extinction curve of Calar Alto for Run 3. Its effect is negligible in $Mg_2$ measurements, and very small in $D_{4000}$ (less than 1%). To correct for interstellar extinction we have used the averaged interstellar extinction curve of Savage and Mathis (1979), and the reddenings derived by Burstein & Heiles (1982). Again $Mg_2$ does not change and $D_{4000}$ is only affected at a $\sim$ 1% level. The exception is PKS 0745−191, which is located near the galactic plane. Using the colour excess of Romanishin (1987) (E(B-V)$\simeq$ 0.38) the correction for this galaxy is $\Delta D_{4000} \simeq 0.20$).

Flux calibration of the spectra has been achieved using exposures of Oke (1974) and Massey et al. (1988) standards (1, 3 and 7 stars for Runs 1, 2 and 3 respectively). All the calibration curves of each run were averaged and the flux calibration errors were estimated by the differences among the indices measured with different curves (typically $\Delta Mg_2 \simeq 0.006$, $\Delta D_{4000} \simeq 0.03$). To check that our $Mg_2$ indices are in the same system as that of the Lick group we measured line-strengths in a sample of G–K giant stars in common with Faber et al. (1985). The comparison is presented in Table 3 and plotted in Figure 3. In average, our $Mg_2$ indices are systematically $\sim$ 0.014 magnitudes below the Lick values, with a standard deviation around this offset of 0.006 mag., in agreement with our internal errors. It should be noted that the Lick measurements are flux calibrated using a tungsten lamp and there is no *a priori* reason to expect that our values are on the same system as theirs. In fact, our offset is similar to those found by other authors [Davies, Sadler & Peletier (1993) and González (1993) reported an offset of 0.017 mag., and Carollo & Danziger (1994) found 0.010 mag.], being their and our measurements in a *true* photometric system. The $Mg_2$ values listed in the tables of the next sections have not been corrected for that offset (i.e., we have kept them in our system).

The error propagation through the data reduction was followed by creating, from photon and readout noises, auxiliary error images which were reduced in parallel to the galaxy frames. To estimate the effects of these errors we measured spectral indices in galaxy spectra in which Gaussian noise, given by the error images, was added. These photon statistics errors were added quadratically to the error due to the flux calibration and, in the case of Run 3, to flatfielding uncertainties. These two last errors usually dominated in the central parts of the galaxies, whereas photon and readout noise were the principal sources of uncertainty in the outer regions. Once the errors in each spatial increment were known, the spectra were binned in the spatial direction with the aim of maintaining a minimum value of the signal-to-noise which guaranteed a reasonable maximum error in the index measurements (typically $\Delta Mg_2 \approx 0.030$, $\Delta D_{4000} \approx 0.20$).

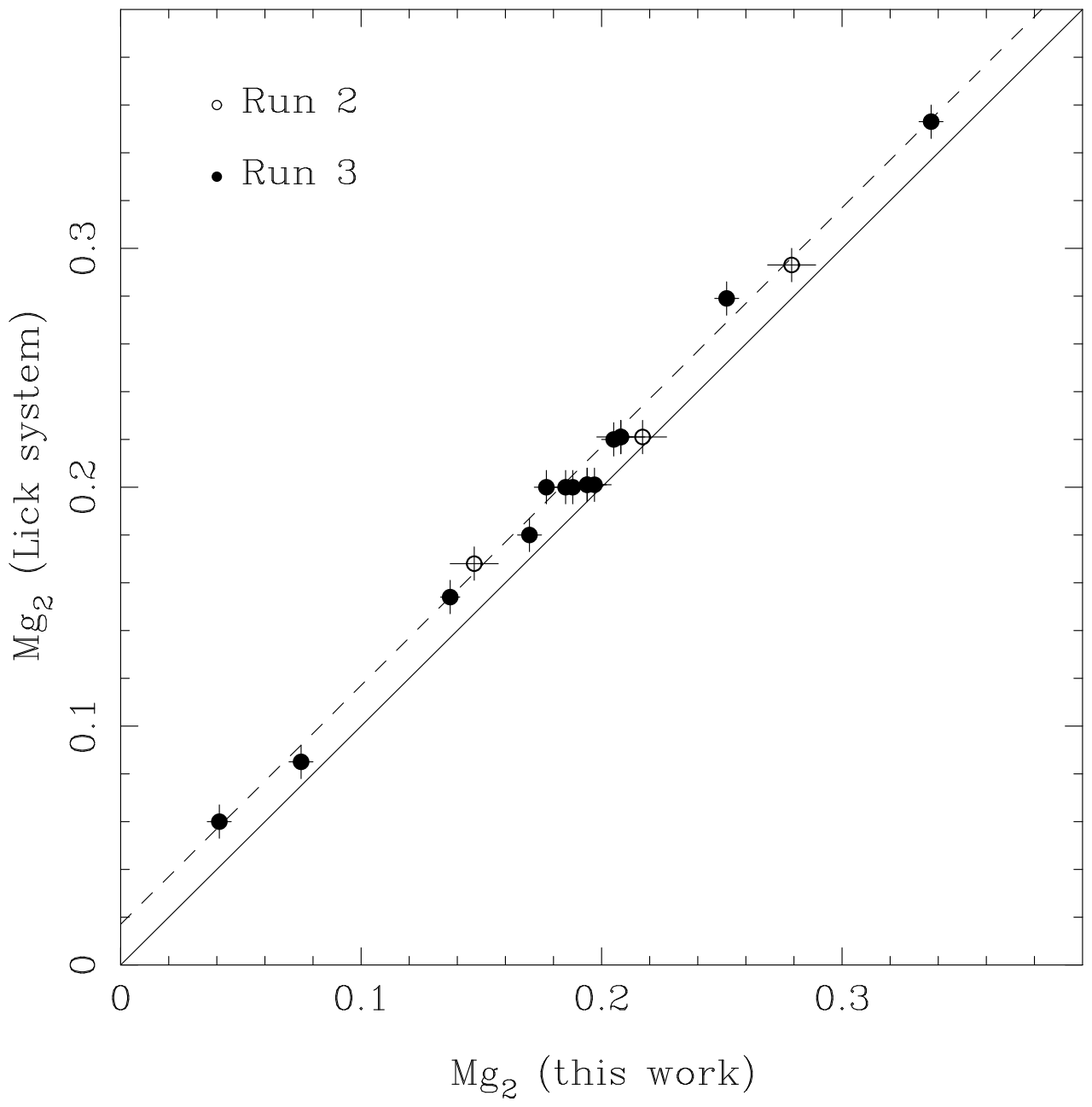

**Figure 3.** Comparison of our $Mg_2$ indices in a sample of G–K giant stars (Table 3) with those measured by Faber et al. (1985). The systematic offset (0.014 magnitudes) is shown by the dashed line. Different symbols have been employed for different observing runs.

**Table 3.** $Mg_2$ measurements in giant stars.

| Star | Run | $Mg_2^{Lick}$ | $Mg_2$ | $Mg_2^{Lick}-Mg_2$ |
|---|---|---|---|---|
| HR1015 | 2 | 0.293 | $0.279 \pm 0.010$ | 0.014 |
| HR3360 | 3 | 0.279 | $0.252 \pm 0.005$ | 0.027 |
| HR3427 | 2 | 0.168 | $0.147 \pm 0.010$ | 0.021 |
| HR5888 | 3 | 0.154 | $0.137 \pm 0.004$ | 0.017 |
| HR6817 | 3 | 0.200 | $0.188 \pm 0.005$ | 0.012 |
| " | 3 | " | $0.177 \pm 0.005$ | 0.023 |
| " | 3 | " | $0.185 \pm 0.005$ | 0.015 |
| HR7148 | 3 | 0.220 | $0.205 \pm 0.005$ | 0.015 |
| HR7317 | 3 | 0.353 | $0.337 \pm 0.005$ | 0.016 |
| HR8165 | 2 | 0.221 | $0.217 \pm 0.010$ | 0.004 |
| " | 2 | " | $0.208 \pm 0.010$ | 0.013 |
| " | 3 | " | $0.208 \pm 0.005$ | 0.013 |
| HD7010 | 3 | 0.180 | $0.170 \pm 0.005$ | 0.010 |
| HD11004 | 3 | 0.085 | $0.075 \pm 0.005$ | 0.010 |
| HD165195 | 3 | 0.060 | $0.041 \pm 0.005$ | 0.019 |
| HD199580 | 2 | 0.201 | $0.194 \pm 0.010$ | 0.007 |
| " | 3 | " | $0.194 \pm 0.005$ | 0.007 |
| " | 3 | " | $0.197 \pm 0.005$ | 0.004 |

## 3 CENTRAL AND GLOBAL INDICES

### 3.1 The data

In this section we present measurements of the central and global $Mg_2$ and $D_{4000}$ indices in the galaxies of the sample. It must be noted that if a fixed angular aperture were employed for all the galaxies, the observed values would be affected, in a different proportion depending on the cluster redshift, by the spectral gradients. Therefore, it is necessary to use distance-independent indices to compare the measurements for different galaxies. As a first step, we have measured central spectral indices ($Mg_2^D$, $D_{4000}^D$) employing a standard metric aperture size that corresponds to $4''$ projected at the distance of the Coma cluster. This allows us to compare our $Mg_2$ indices with those obtained by Davies et al. (1987) for a large sample of elliptical galaxies. The results, together with their errors and some galaxy parameters, are listed in Table 4.

Other distance-independent indices, which account for the different sizes of the galaxies, have been defined simulating circular apertures related to the effective radius of each object. These circular apertures are simulated adding radius-weighted spectra at different distances from the centres. In this way we define "central" indices corresponding to an aperture of $r_e/10$ diameter centred at the nucleus ($Mg_2^c$, $D_{4000}^c$) and "global" indices for an aperture of $r_e$ diameter ($Mg_2^g$, $D_{4000}^g$), where $r_e$ is the effective radius from Schombert (1986, 1987), and Romanishin & Hintzen (1988). The $r_e$ for the central galaxy in A 2319 has been estimated from a comparison with galaxies of similar characteristics. These central and global indices, listed with their errors in Table 4, were not measured for PKS 0745−191 and A 644 due to their poor signal-to-noise spectra outside the centre and to the lack of photometric data in the literature. Columns (6) and (9) of the table list the central and global mean radii, weighted by light and distance, used to mimic the circular apertures.

### 3.2 Modelling star formation in cooling flow galaxies

In order to interpret the observed spectral properties of the galaxies, we have used an evolutionary stellar population synthesis technique. The aim is to predict the changes produced in the spectra of an old stellar populations (elliptical galaxy) when a continuous star formation, induced by the cooling-flow gas, is taking place. We have closely followed the procedure described in GEA (their section 3.3), so the details will not be discussed here. The only difference is that we have now used the Bruzual & Charlot (1993) models instead of the ones mentioned in GEA, although in the wavelength region considered here both sets of models give very similar results. For a more detailed discussion of the available galaxy evolution models, see Cole et al. (1994).

In brief, the Bruzual & Charlot (1993) models were used to produce synthetic spectra for a stellar population with a given IMF and star formation history, which we superimpose on the spectrum of a normal giant elliptical with $Mg_2 = 0.340$ and $D_{4000} = 2.40$. We computed models with different IMF's (Salpeter 1955, Miller & Scalo 1979, and Scalo 1986). In all cases we truncated the IMF below $0.1\,M_\odot$. Lowering this value has no effect on the modeled spectral properties, but alters the $M/L$ ratio of the accretion population. Constant SFR's of different durations (5–10 Gyr) were considered. The models can be parameterised by the fraction $f_V$ of the V light that comes from the accretion population. In turn, this can be related to the SFR using the expression:

$$\mathrm{SFR} = \frac{f_V L_V (M/L_V)_{AP}}{t} \quad (1)$$

where $L_V$ is the V luminosity of the galaxy, $(M/L_V)_{AP}$ is the luminous mass-to-light ratio of the accretion population,

**Table 4.** Central and global indices.

| Cluster | $\dot{M}$ ($M_\odot$/yr) | $r_e$ (arcsec) | $Mg_2^D$ | $D_{4000}^D$ | $r^c$ (arcsec) | $Mg_2^c$ | $D_{4000}^c$ | $r^g$ (arcsec) | $Mg_2^g$ | $D_{4000}^g$ |
|---|---|---|---|---|---|---|---|---|---|---|
| A 262 | 47 | 72.8 | 0.316 | 2.34 | 2.20 | 0.313 | 2.33 | 18.21 | 0.288 | 2.24 |
| | | | 0.009 | 0.04 | | 0.009 | 0.04 | | 0.011 | 0.06 |
| 2A 0335+096 | 142 | 39.0 | 0.279 | 2.08 | 1.24 | 0.276 | 2.12 | 9.03 | 0.155 | 1.88 |
| | | | 0.009 | 0.04 | | 0.009 | 0.04 | | 0.017 | 0.09 |
| PKS 0745−191 | 702 | — | 0.116 | 1.31 | — | — | — | — | — | — |
| | | | 0.045 | 0.04 | | — | — | | — | — |
| A 644 | 326 | — | 0.263 | 2.44 | — | — | — | — | — | — |
| | | | 0.021 | 0.10 | | — | — | | — | — |
| A 1126 | 222 | 18.3 | 0.280 | 1.82 | 0.61[a] | 0.284 | 1.81 | 4.99[a] | 0.257 | 1.74 |
| | | | 0.009 | 0.03 | | 0.008 | 0.01 | | 0.008 | 0.02 |
| A 1795 | 478 | 34.4 | 0.209 | 1.61 | 1.45 | 0.209 | 1.61 | 8.29 | 0.189 | 1.60 |
| | | | 0.013 | 0.03 | | 0.013 | 0.03 | | 0.011 | 0.03 |
| A 2124 | 0 | 33.3 | 0.282 | 2.14 | 1.43 | 0.275 | 2.12 | 8.31 | 0.221 | 1.94 |
| | | | 0.010 | 0.04 | | 0.009 | 0.04 | | 0.010 | 0.04 |
| A 2199 | 150 | 34.7 | 0.299 | 2.09 | 1.49 | 0.300 | 2.09 | 9.16 | 0.266 | 1.94 |
| | | | 0.008 | 0.06 | | 0.008 | 0.06 | | 0.008 | 0.06 |
| A 2319 | 66 | 40.5[b] | 0.297 | 2.02 | 1.52 | 0.293 | 2.03 | 10.58 | 0.255 | 1.84 |
| | | | 0.010 | 0.05 | | 0.010 | 0.04 | | 0.010 | 0.04 |
| A 2634 | 0 | 13.8 | 0.286 | 2.21 | 0.48 | 0.295 | 2.21 | 3.66 | 0.270 | 2.11 |
| | | | 0.007 | 0.02 | | 0.009 | 0.03 | | 0.007 | 0.02 |

Errors are given underneath the data. Mass accretion rates are from Arnaud (1988), Heckman et al. (1989), and Edge et al. (1992).
[a] Radii listed for this galaxy correspond to the CCD spectral observations and are only applied to the $Mg_2$ index. Corresponding radii for $D_{4000}$ (from IPCS observations) are $r^c = 0.89$ and $r^g = 5.20$.
[b] The effective radius for A 2319 has been estimated from a comparison with galaxies of similar characteristics.

given by the models, and $t$ is the time over which the star formation has taken place.

The model results will be presented and compared with the data in the next section. But before that, a word of caution on the use of this kind of models to interpret the observations. There are still important uncertainties in the building of evolutionary models (cf. Cole et al. 1994). In this work, the main uncertainty comes from the fact that the stellar population models that we have used have solar metallicity and solar abundance ratios. Since $Mg_2$ and, probably, $D_{4000}$ depend on the metal abundance (see, e.g., Worthey 1994), and the cooling IGM has metallicities that could be below solar (Mushotzky 1992), the absolute values of the model indices could be affected by systematics, and have to be taken with a pinch of salt. However, we are confident that the *relative changes* and *trends* that the new stars produce in the galaxy spectral indices are reasonably well represented by the models.

### 3.3 Discussion of central and global indices

Now we discuss the possible effects of the cooling flows in the indices measured in the galaxies of the sample. As usual, the importance of the cooling flow is estimated from the mass accretion rates (in $M_\odot$/year) inferred from X-ray observations. For the galaxies of our sample, these are listed in Table 4, where the values from Edge et al. (1992) have been used when available⋆. It must be noted that errors associated with these rates are usually very high (see, e.g., Fabian et al. 1991). For A 1126 we have used the mass flow rate quoted by Heckman et al. (1989). Cooling times of the gas in A 2124 and A 2634 (from Arnaud 1988) are well above the Hubble time and therefore a cooling flow can not exist.

In order to investigate the possible effects of the cooling flows in the measured indices, we plot in Figure 4 the central and global $Mg_2$ and $D_{4000}$ indices against the mass accretion rates. In the case of the central $Mg_2$ we have included the cD's of the sample of GEA with measured accretion rates. Note that the brightest *normal* elliptical galaxies attain $Mg_2$ values around 0.300–0.350 mag, with an intrinsic scatter of 0.025 mag. When the central indices (using the Davies et al. 1987 aperture) are considered, there is a clear trend in the sense that both $Mg_2^D$ and $D_{4000}^D$ are smaller when the mass flow increases. An exception to this trend is A 644 for which an anomalously high $D_{4000}$ has been measured. No explanation for this value has been found.

A similar correlation for the 4000 Å break was previously presented by JFN. However, they used a fixed angular aperture and, since in their sample $\dot{M}$ and $D_{4000}$ were both highly correlated with redshift, the reported correlation could be due to an aperture effect (the correlation of $\dot{M}$ with redshift can be explained as a selection effect, since high mass accretion rate cooling flows are more likely to be found at high distances, whereas gradients in $D_{4000}$ could account for the correlation between index and redshift). Since our indices are distance-independent we can confirm that the correlation of JFN is real and that galaxies which host prominent cooling flows tend to exhibit signatures of star formation in their spectra (see Section 4.2).

It is clear from Figure 4 that the correlation with $\dot{M}$ is somewhat weakened when the global indices, corresponding to an important fraction of the galaxy, are considered. This can be understood since global indices are influenced by the mass deposition profile along the galaxy and by the intrinsic spectral gradients, and both of them could significantly change from galaxy to galaxy (see e.g. McNamara & O'Connell 1992, and GEA). The combination of these two effects makes global spectral indices in cooling-flow galaxies quite difficult to interpret (Section 4.2). However, we must stress that the correlations with $\dot{M}$, particularly that of $D_{4000}^g$, are still strong.

⋆ $H_0 = 50\,\mathrm{km\,s^{-1}Mpc^{-1}}$ has been used throughout the paper.

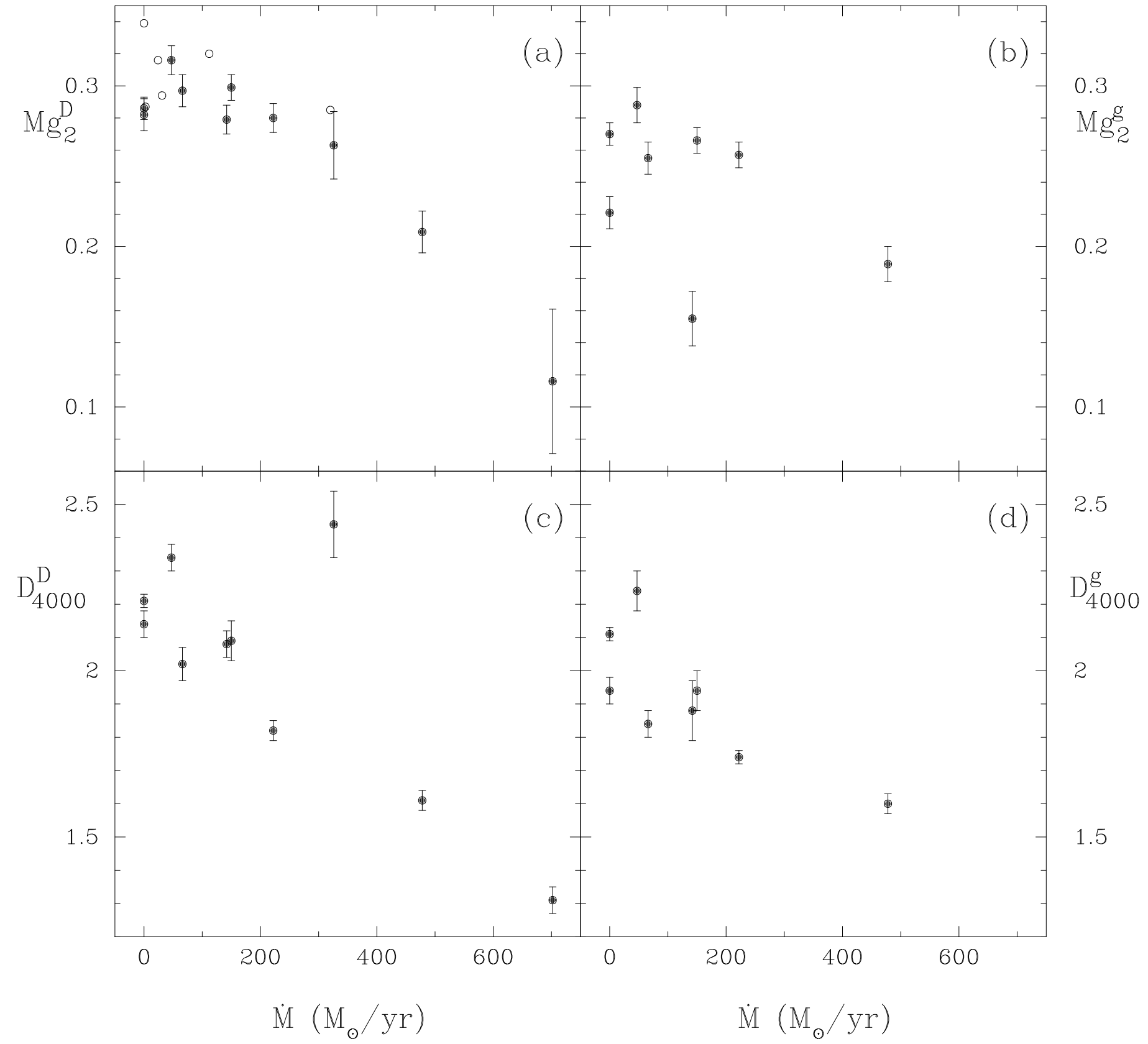

**Figure 4.** Central ($Mg_2^D$, $D_{4000}^D$) and global ($Mg_2^g$, $D_{4000}^g$) indices versus mass accretion rate. The central indices refer to those measured in the Davies et al. (1987) aperture. Open circles in (a) correspond to the galaxies in GEA with measured accretion rates (A 957, Sersic 40/6, A 754, NGC 4696, A 496 and PKS 2354-35, in order of increasing $\dot{M}$). The deviant point in (c) corresponds to A 644. There is a clear correlation between the measured central indices and $\dot{M}$ ((a) and (c)). Global indices ((b) and (d)) are influenced by the intrinsic spectral gradients and the correlation is poorer.

The comparison of these results with the models introduced in Section 3.2 allows an interpretation of the spectral indices in terms of star formation from the accreted gas. In order to simultaneously study the variations in $Mg_2$ and $D_{4000}$ we present in Figure 5 an index-index plot. The central indices of the sample galaxies are plotted as crosses, whose length is given by the errors, and open circles, being their sizes proportional to the mass accretion rates. We also show the locus of the normal elliptical galaxies, where $D_{4000}$ values are taken from Hamilton (1985) and Munn (1992), whereas $Mg_2$ indices come from Davies et al. (1987). These $Mg_2$ indices have been transformed to our photometric system (see Section 2.2). The straight line is a least-square fit to these data.

It is apparent from this Figure that cooling-flow galaxies fall outside the normal ellipticals' region, increasing the deviations with the gas flow (as it was expected from Figure 4). In order to compare these indices with the predictions of the stellar populations models, we plot the expected indices for a giant elliptical galaxy in which star formation has taken place at a constant rate over the last 5 and 10 Gyrs. Small dots in the model lines correspond to a fixed fraction $f_V$ of the V light of the galaxy that comes from this accretion population. It is clear that these predictions do not depend very much on the choice of IMF or the time scale of the star formation (provided that it is a significant fraction of the age of the Universe) and, in any case, cooling flow galaxies do follow the trends of the models. This implies that the values that these galaxies attain in both indices are roughly consistent with a constant star formation in these objects.

In order to confirm that the accretion population increases in importance with the mass accretion rates, we have estimated the luminosity fractions $f_V$ interpolating the indices in the models corresponding to a Miller & Scalo (1979) IMF for a constant SFR during 5 Gyr. Note that the results are not very sensitive to the choice of IMF. These luminosity fractions can then be converted to actual star formation rates using equation (1). We have used a constant value $M_V \simeq -24$ for the total absolute magnitude of the galaxies (Schombert 1986, 1987), since the brightest cluster galaxies are reasonably good standard candles. For this IMF and $t = 5$ Gyr, we can estimate the actual SFR assuming that the fraction of V light due to the accretion population does not vary with radius. This is of course an oversimplification, but no better approximation can be done until we know the radial profile of the mass deposition rate (Section 4.2). The deduced SFR's, from the central $Mg_2$ and $D_{4000}$ indices (filled and open circles respectively) are plotted against the estimated mass accretion rates in Figure 6. This confirms that there is a *strong* correlation and that the actual SFR's deduced from both indices are consistent given the uncertainties inherent to the models.

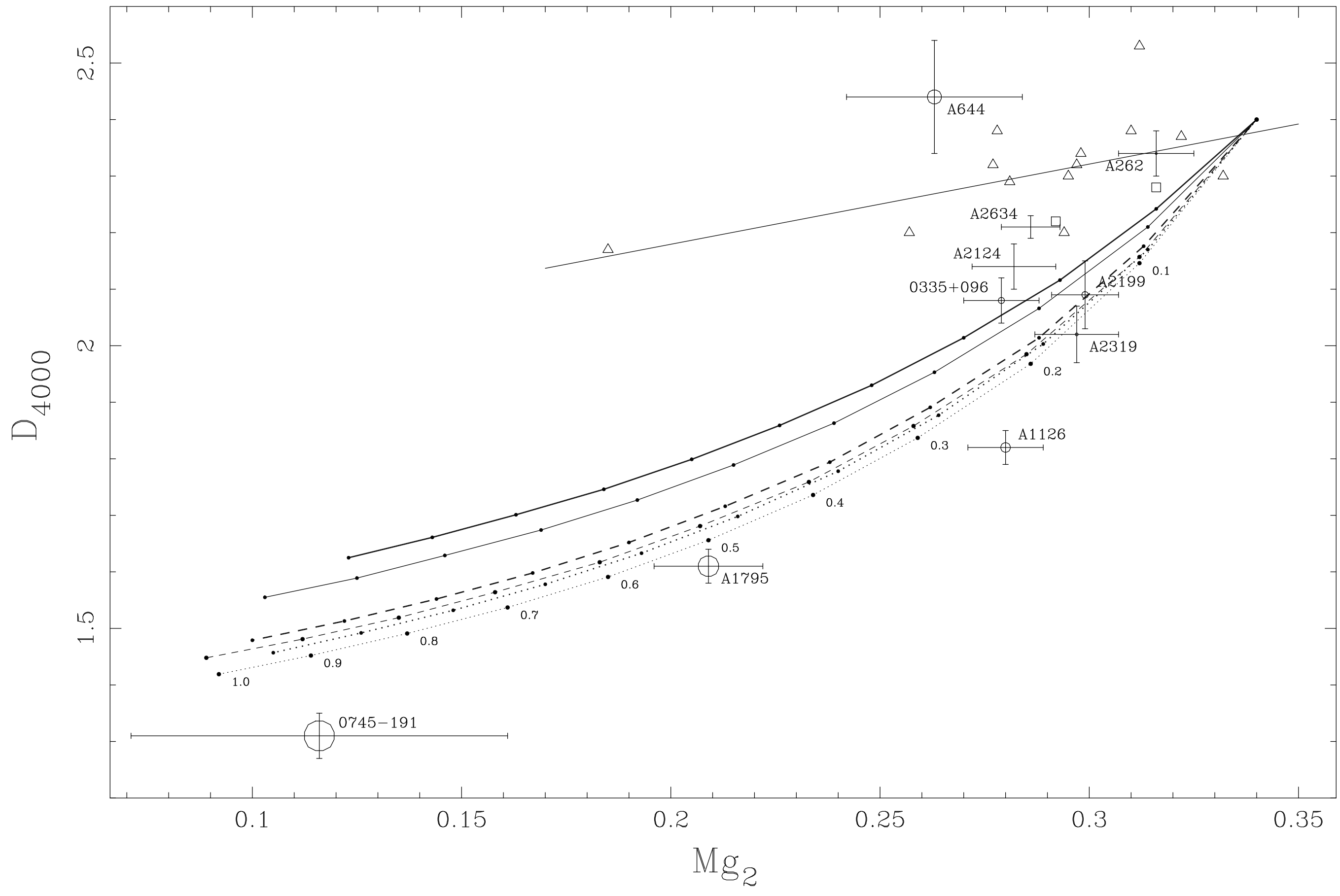

**Figure 5.** $Mg_2$–$D_{4000}$ diagram in which the spectral indices of the sample galaxies are plotted as open circles and crosses, with lengths given by the errors. The circle sizes are proportional to the mass accretion rate of the cluster. The rest of the open symbols represent normal elliptical galaxies from Hamilton (1985) (triangles) and Munn (1992) (squares). The straight line is a least-squares fit to these data. Curves show the predictions of stellar population models with different IMF's: Scalo (1986) (full lines), Miller & Scalo (1979) (dashed lines) and Salpeter (1955) (dotted lines). Models assume that a constant star formation has been taken place over the last 5 (thin line) and 10 Gyr (thick line). Small dots in the model lines correspond to a fixed fraction $f_V$ (indicated by the numbers) of the V light that comes from the accretion population.

Interestingly, the ratio between the estimated SFR's and $\dot{M}$ is quite similar from galaxy to galaxy, with an average value $\langle \mathrm{SFR}/\dot{M} \rangle \simeq 17\%$. If a time of 10 Gyr for the mass deposition were assumed we would obtain $\langle \mathrm{SFR}/\dot{M} \rangle \simeq 5\%$. In any case, we obtain relatively small fractions of the X-ray inferred gas flow. However, the derived fraction is a *lower limit* for two reasons: first, the computed SFR includes only stars with $M > 0.1 M_\odot$, and, depending on uncertain extrapolations of the IMF towards lower masses, we could be underestimating the SFR by a factor of a few. Second, the $\dot{M}$ inferred from X-ray observations is the *total* accretion rate. If, as the X-ray data suggest (see, e.g., Fabian 1994), a large fraction of the mass is deposited at radii that are large compared with the optical radius of the galaxies, the amount of accreting gas available for star formation *within* the galaxy (i.e., the region contributing to $L_V$) would be substatially reduced. In this respect, the fact that SFR/$\dot{M}$ is quite constant from galaxy to galaxy seems to indicate that the star formation efficiency of the accreting gas remains fairly constant over almost 3 orders of magnitud in $\dot{M}$ and very different physical conditions. That would be quite puzzling unless the actual SFR were a substantial fraction (perhaps all) of the mass accreted within the "optical galaxy", and the mass deposition profile is reasonably similar for all the clusters. Note also that if the cooling flows have persisted for several Gyr, the accreted population can account, in some cases, for a very significant fraction of the luminous matter of the galaxies. †

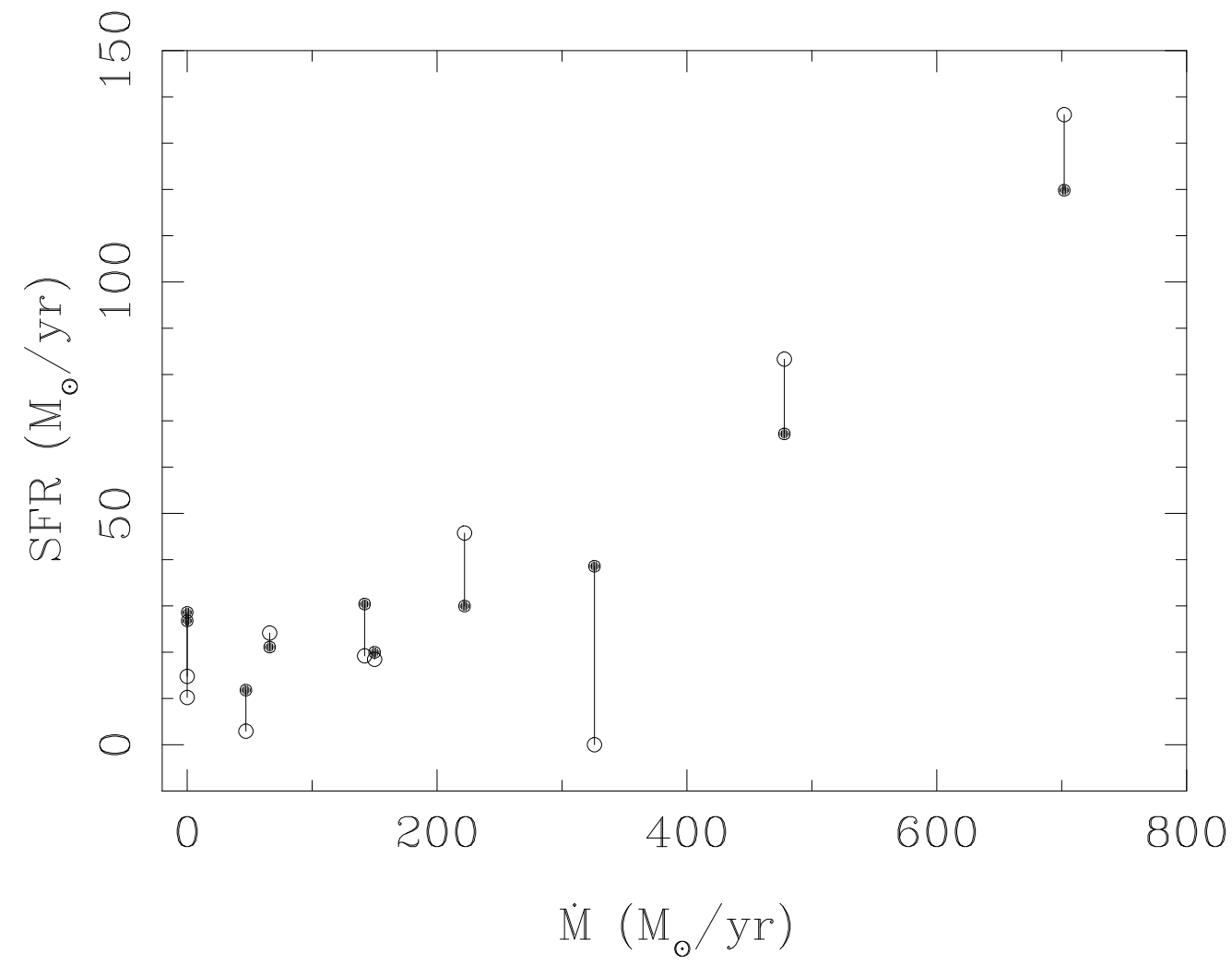

**Figure 6.** Model-derived SFR's vs mass accretion rates. Filled and open symbols correspond to $f_V$ interpolated in the model of Figure 5 (Miller & Scalo 1979 IMF and $t = 5$ Gyr) from measured $Mg_2$ and $D_{4000}$ respectively. For each galaxy, both estimates of SFR have been joined by a straigth line.

## 4 INDEX GRADIENTS

### 4.1 The data

Our long-slit observations allow the extraction of spectra at different galactocentric distances. Prior to this, the spectra were shifted, by a fraction of a pixel in the spatial direction, in order to obtain symmetrical brightness profiles at both sides of the galaxy centre. Furthermore, several spatial increments in the outer parts were co-added as explained in Section 2. The variations of $Mg_2$ and $D_{4000}$ with radius and their formal errors are listed in Table 5 for the eight galaxies of the sample with appropriate exposures. The table shows that the $D_{4000}$ index can usually be measured at distances further out from the centre, due to its wider spectral bands. For some objects, a secondary nucleus falls into the slit. In that case, a superscript "b" has been employed to indicate the radii where the brightness of these nuclei dominates over the main galaxy profile. Analogously, for A 1126, a companion galaxy was observed in the slit (denoted by "c"). For A 1126 the table lists measurements both with the IPCS II (noted by †) and the CCD detectors.

The radial gradients in $Mg_2$ and $D_{4000}$ are plotted in Figures 7a–7h in a logarithmic scale. The figures show that indices measured at both sides of the galaxy usually agree within the errors. With the exception of A 262 and 2A 0335+096, gradients were traced out to $r \sim r_e$ for both $Mg_2$ and $D_{4000}$. Due to its poorly determined gradients (low signal-to-noise in the outer parts due to clouds), 2A 0335+096 has been excluded from the subsequent analysis. It is well known that $Mg_2$ indices within normal elliptical galaxies follow a linear relation with $\log r$ (e.g. GEA; Davies et al. 1993; González 1993). However, in some galaxies of our sample (e.g. A 1795 and A 2124) we find a significant change in slope at a intermediate radius, with a small gradient in the inner regions and the indices falling dramatically in the outer parts. We find that, in general, gradients vary markedly from galaxy to galaxy (e.g. compare A 1126 and A 2199). This diversity of behaviours is similar to that reported for elliptical galaxies.

In order to quantify the gradients and comparing them with those found in elliptical galaxies, we computed linear regression fits to the data in the logarithmic representation of Figures 7. These fits, computed with error-weighting, and excluding points with either $r < 1.5''$ (usually affected by seeing) or corresponding to secondary nuclei (open symbols), are plotted as full lines in the figures. The derived gradients ($G_{\mathrm{Mg}_2} = d\mathrm{Mg}_2/d\,log_{10}(r)$, $G_{D_{4000}} = dD_{4000}/d\,log_{10}(r)$) and their formal errors are listed in Table 6. With the aim of studying the slope changes observed for some objects, we have also computed gradients to the inner regions of the galaxies ($1.5'' < r \leq r_e/2$). The upper limit was chosen visually as the radius where the slope changes occur. These internal gradients, denoted as $G^{\mathrm{int}}_{\mathrm{Mg}_2}$ and $G^{\mathrm{int}}_{D_{4000}}$, and their errors are also given in Table 6. They are plotted in Figures 7 as dashed lines. The dashed lines beyond $r_e/2$ correspond to linear fits using the outer points and forcing them to join with the internal gradients at $r = r_e/2$. Note that in some cases (A 2634) these outer gradients are quite uncertain due to the small number of points in the outer parts. This separation in inner and outer gradients allows to visualise the slope changes mentioned above.

### 4.2 Discussion of spectral gradients

The error-weighted mean of the $Mg_2$ gradients for the galaxies of the sample is $\langle G_{\mathrm{Mg}_2} \rangle = -0.065 \pm 0.027$ ‡. If only cooling flow galaxies are considered, we obtain a mean gradient of $-0.055 \pm 0.019$. The mean $Mg_2$ gradients found in different samples of early-type galaxies are $-0.058 \pm 0.027$ (GEA), $-0.059 \pm 0.022$ (Davies et al. 1993) and $-0.062 \pm 0.024$ (Carollo, Danziger & Buson 1993). Therefore, we find a mean $Mg_2$ gradient in good agreement with that found in normal elliptical galaxies. Due to the slope changes at intermediate radius, mean internal gradients are lower than those computed above, being $\langle G^{\mathrm{int}}_{\mathrm{Mg}_2} \rangle = -0.055 \pm 0.025$ for the whole sample and $-0.045 \pm 0.014$ for the galaxies with cooling flows. The main conclusion is that inner $Mg_2$ gradients for cooling flow galaxies are, on average, flatter than those of normal ellipticals. Applying a two–sample $t$ statistic, the confidence level of this statement is ∼90% (see Figure 8 below). GEA measured internal $Mg_2$ gradients in three cD galaxies. Their cooling flow galaxy (PKS 2354−35) also possessed a shallow gradient, although the results presented here are of a much better quality.

We have also computed mean $D_{4000}$ gradients, obtaining $\langle G_{D_{4000}} \rangle = -0.30 \pm 0.14$ (whole sample) and $-0.28 \pm 0.14$ (cooling flow galaxies). In this case, the mean internal gradients are $\langle G^{\mathrm{int}}_{D_{4000}} \rangle = -0.19 \pm 0.13$ and $-0.16 \pm 0.14$ for the whole sample and the galaxies with cooling flows respectively. The flattening of gradients in the inner regions is evident for this index. The only studies of $D_{4000}$ gradients in elliptical galaxies are those of Munn (1992) and Davidge & Clark (1994). Munn presented gradients for 7 early-type galaxies, two of which are classified as ring galaxies. Excluding these two objects, the mean $D_{4000}$ gradient is $-0.32 \pm 0.04$ (Note the low dispersion of data around the

† The SFR/$\dot{M}$ ratio is almost intependent on the Hubble constant: SFR$\propto H_0^{-2}$ and $\dot{M} \propto H_0^{-\gamma}$, where $\gamma = 1.7$–$2$, depending on the mass deposition profile (Edge 1994, private communication).

‡ The quoted errors are the *r.m.s.* dispersions around the mean values. The statistical error in the mean would be $r.m.s./\sqrt{n}$ with $n$ the number of points.

**Table 5.** Line-strength indices along radius.

**Abell 262**

| r(″) | $Mg_2$ | $\Delta Mg_2$ | r(″) | $D_{4000}$ | $\Delta D_{4000}$ |
|---|---|---|---|---|---|
| | | | -28.0 | 2.10 | 0.16 |
| | | | -24.1 | 2.10 | 0.16 |
| | | | -21.6 | 2.18 | 0.17 |
| -27.0 | 0.260 | 0.032 | -19.8 | 2.21 | 0.17 |
| -21.8 | 0.263 | 0.034 | -18.3 | 2.13 | 0.16 |
| -18.8 | 0.235 | 0.031 | -17.1 | 2.31 | 0.19 |
| -16.8 | 0.293 | 0.034 | -16.1 | 2.15 | 0.16 |
| -15.2 | 0.292 | 0.031 | -15.1 | 2.27 | 0.16 |
| -14.1 | 0.353 | 0.034 | -14.1 | 2.29 | 0.15 |
| -13.1 | 0.271 | 0.030 | -13.1 | 2.22 | 0.12 |
| -12.1 | 0.254 | 0.027 | -12.1 | 2.15 | 0.10 |
| -11.1 | 0.277 | 0.024 | -11.1 | 2.24 | 0.10 |
| -10.1 | 0.283 | 0.022 | -10.1 | 2.25 | 0.09 |
| -9.1 | 0.281 | 0.020 | -9.1 | 2.18 | 0.07 |
| -8.0 | 0.313 | 0.018 | -8.0 | 2.24 | 0.07 |
| -7.0 | 0.313 | 0.016 | -7.0 | 2.22 | 0.06 |
| -6.0 | 0.306 | 0.014 | -6.0 | 2.29 | 0.06 |
| -5.0 | 0.316 | 0.012 | -5.0 | 2.25 | 0.05 |
| -4.0 | 0.321 | 0.011 | -4.0 | 2.31 | 0.05 |
| -3.0[b] | 0.306 | 0.011 | -3.0[b] | 2.36 | 0.05 |
| -2.0 | 0.319 | 0.011 | -2.0 | 2.34 | 0.05 |
| -1.0 | 0.305 | 0.011 | -1.0 | 2.31 | 0.05 |
| 0.3 | 0.320 | 0.011 | 0.3 | 2.31 | 0.04 |
| 1.0 | 0.311 | 0.011 | 1.0 | 2.36 | 0.05 |
| 2.0 | 0.319 | 0.012 | 2.0 | 2.38 | 0.05 |
| 3.0 | 0.318 | 0.014 | 3.0 | 2.31 | 0.06 |
| 4.0 | 0.273 | 0.015 | 4.0 | 2.25 | 0.06 |
| 5.0 | 0.310 | 0.016 | 5.0 | 2.17 | 0.06 |
| 6.0 | 0.284 | 0.020 | 6.0 | 2.19 | 0.08 |
| 7.0 | 0.294 | 0.022 | 7.0 | 2.16 | 0.09 |
| 8.0 | 0.297 | 0.023 | 8.0 | 2.37 | 0.11 |
| 9.1 | 0.277 | 0.025 | 9.1 | 2.25 | 0.11 |
| 10.1 | 0.251 | 0.029 | 10.1 | 2.02 | 0.11 |
| 11.1 | 0.287 | 0.031 | 11.1 | 2.13 | 0.13 |
| 12.2 | 0.333 | 0.031 | 12.1 | 2.18 | 0.15 |
| 13.7 | 0.352 | 0.032 | 13.1 | 1.99 | 0.15 |
| 15.6 | 0.305 | 0.034 | 14.1 | 2.18 | 0.16 |
| 18.1 | 0.327 | 0.033 | 15.2 | 2.34 | 0.18 |
| 21.8 | 0.296 | 0.034 | 16.6 | 2.51 | 0.24 |
| 28.5 | 0.295 | 0.034 | 18.3 | 2.43 | 0.21 |
| | | | 20.3 | 2.22 | 0.18 |
| | | | 22.8 | 2.22 | 0.19 |
| | | | 28.8 | 2.33 | 0.22 |

**2A 0335+096**

| r(″) | $Mg_2$ | $\Delta Mg_2$ | r(″) | $D_{4000}$ | $\Delta D_{4000}$ |
|---|---|---|---|---|---|
| | | | -8.0 | 2.22 | 0.20 |
| -6.5 | 0.200 | 0.029 | -5.2 | 1.97 | 0.17 |
| -4.2 | 0.261 | 0.029 | -4.0 | 2.03 | 0.14 |
| -3.0 | 0.259 | 0.023 | -3.0 | 2.00 | 0.11 |
| -2.0 | 0.244 | 0.018 | -2.0 | 2.11 | 0.09 |
| -1.0 | 0.281 | 0.015 | -1.0 | 2.07 | 0.06 |
| 0.3 | 0.274 | 0.014 | 0.3 | 2.11 | 0.06 |
| 1.0 | 0.283 | 0.015 | 1.0 | 2.07 | 0.06 |
| 2.0 | 0.277 | 0.019 | 2.0 | 2.17 | 0.09 |
| 3.0 | 0.272 | 0.026 | 3.0 | 2.16 | 0.13 |
| 4.2 | 0.244 | 0.027 | 4.2 | 2.14 | 0.14 |
| 5.4[b] | 0.286 | 0.024 | 5.4[b] | 2.14 | 0.13 |
| 6.4[b] | 0.269 | 0.020 | 6.4[b] | 2.21 | 0.12 |
| 7.4[b] | 0.250 | 0.021 | 7.4[b] | 2.24 | 0.13 |
| 8.7[b] | 0.212 | 0.029 | 8.4[b] | 2.32 | 0.21 |

**Abell 1126**

| r(″) | $Mg_2$ | $\Delta Mg_2$ | r(″) | $D_{4000}$ | $\Delta D_{4000}$ |
|---|---|---|---|---|---|
| -11.6 | 0.201 | 0.029 | | | |
| -9.6[†] | 0.232 | 0.026 | | | |
| -9.0 | 0.225 | 0.028 | -21.0[†] | 1.59 | 0.15 |
| -7.8[b] | 0.199 | 0.022 | -16.2[†] | 1.37 | 0.13 |
| -6.8[b] | 0.208 | 0.015 | -14.4[†] | 1.76 | 0.21 |
| -6.0[†b] | 0.242 | 0.025 | -13.2[†] | 1.85 | 0.18 |
| -5.8[b] | 0.263 | 0.011 | -12.0[†] | 1.55 | 0.11 |
| -4.9[b] | 0.277 | 0.012 | -10.8[†] | 1.58 | 0.09 |
| -4.8[†b] | 0.238 | 0.022 | -9.6[†] | 1.67 | 0.07 |
| -4.1[b] | 0.238 | 0.011 | -8.4[†] | 1.78 | 0.05 |
| -3.6[†b] | 0.266 | 0.023 | -7.2[†b] | 1.81 | 0.04 |
| -3.1 | 0.250 | 0.013 | -6.0[†b] | 1.83 | 0.03 |
| -2.4[†] | 0.263 | 0.023 | -4.8[†b] | 1.85 | 0.02 |
| -2.0 | 0.260 | 0.012 | -3.6[†b] | 1.82 | 0.02 |
| -1.2[†] | 0.258 | 0.022 | -2.4[†] | 1.80 | 0.03 |
| -1.0 | 0.280 | 0.011 | -1.2[†] | 1.80 | 0.02 |
| 0.3[†] | 0.300 | 0.020 | 0.3[†] | 1.82 | 0.02 |
| 0.3 | 0.279 | 0.010 | 1.2[†] | 1.80 | 0.02 |
| 1.0 | 0.275 | 0.011 | 2.4[†] | 1.75 | 0.03 |
| 1.2[†] | 0.287 | 0.022 | 3.6[†] | 1.66 | 0.03 |
| 2.0 | 0.254 | 0.014 | 4.8[†] | 1.69 | 0.04 |
| 2.4[†] | 0.224 | 0.027 | 6.0[†] | 1.75 | 0.06 |
| 3.1 | 0.266 | 0.017 | 7.2[†] | 1.77 | 0.06 |
| 4.1 | 0.243 | 0.018 | 8.4[†] | 1.80 | 0.06 |
| 5.1 | 0.243 | 0.025 | 9.6[†c] | 1.73 | 0.05 |
| 5.4[†] | 0.215 | 0.022 | 10.8[†c] | 1.76 | 0.04 |
| 6.3 | 0.241 | 0.026 | 12.0[†c] | 1.87 | 0.05 |
| 7.7 | 0.271 | 0.029 | 13.2[†c] | 1.82 | 0.07 |
| 9.0 | 0.260 | 0.027 | 14.4[†] | 1.75 | 0.10 |
| 9.6[†c] | 0.274 | 0.028 | 15.6[†] | 1.91 | 0.16 |
| 10.2[c] | 0.316 | 0.024 | 16.8[†] | 1.68 | 0.18 |
| 11.2[c] | 0.262 | 0.019 | 18.6[†] | 1.62 | 0.17 |
| 12.2[c] | 0.299 | 0.024 | 24.0[†] | 1.75 | 0.20 |
| 13.2[†c] | 0.201 | 0.033 | | | |
| 13.8 | 0.238 | 0.029 | | | |

**Abell 1795**

| r(″) | $Mg_2$ | $\Delta Mg_2$ | r(″) | $D_{4000}$ | $\Delta D_{4000}$ |
|---|---|---|---|---|---|
| | | | -30.4 | 1.69 | 0.18 |
| | | | -28.6 | 1.51 | 0.13 |
| | | | -26.9 | 1.42 | 0.11 |
| | | | -25.1 | 1.45 | 0.10 |
| -27.8 | 0.117 | 0.035 | -23.3 | 1.41 | 0.09 |
| -22.4 | 0.162 | 0.033 | -21.5 | 1.55 | 0.09 |
| -19.7 | 0.198 | 0.040 | -19.7 | 1.48 | 0.07 |
| -17.9 | 0.163 | 0.032 | -17.9 | 1.49 | 0.07 |
| -16.1 | 0.166 | 0.029 | -16.1 | 1.52 | 0.06 |
| -14.3 | 0.169 | 0.025 | -14.3 | 1.54 | 0.05 |
| -12.5 | 0.176 | 0.021 | -12.5 | 1.54 | 0.04 |
| -10.7 | 0.198 | 0.019 | -10.7 | 1.64 | 0.04 |
| -9.0 | 0.193 | 0.017 | -9.0 | 1.61 | 0.04 |
| -7.2 | 0.204 | 0.014 | -7.2 | 1.58 | 0.04 |
| -5.4 | 0.198 | 0.012 | -5.4 | 1.59 | 0.03 |
| -3.6 | 0.200 | 0.010 | -3.6 | 1.60 | 0.03 |
| -1.8 | 0.214 | 0.009 | -1.8 | 1.62 | 0.03 |
| 0.4 | 0.209 | 0.009 | 0.4 | 1.61 | 0.03 |
| 1.8 | 0.204 | 0.010 | 1.8 | 1.59 | 0.03 |
| 3.6 | 0.209 | 0.010 | 3.6 | 1.61 | 0.03 |
| 5.4 | 0.186 | 0.011 | 5.4 | 1.59 | 0.03 |
| 7.2 | 0.170 | 0.013 | 7.2 | 1.64 | 0.04 |
| 9.0 | 0.186 | 0.015 | 9.0 | 1.66 | 0.04 |

**Table 5.** – *continued*

**Abell 1795 (contd.)**

| r(″) | $Mg_2$ | $\Delta Mg_2$ | r(″) | $D_{4000}$ | $\Delta D_{4000}$ |
|---|---|---|---|---|---|
| 10.7 | 0.186 | 0.017 | 10.7 | 1.65 | 0.04 |
| 12.5 | 0.187 | 0.021 | 12.5 | 1.66 | 0.05 |
| 14.3 | 0.178 | 0.024 | 14.3 | 1.65 | 0.06 |
| 16.1 | 0.153 | 0.026 | 16.1 | 1.57 | 0.05 |
| 17.9 | 0.129 | 0.027 | 17.9 | 1.53 | 0.06 |
| 19.7 | 0.125 | 0.032 | 19.7 | 1.47 | 0.06 |
| 22.4 | 0.125 | 0.030 | 21.5 | 1.51 | 0.07 |
| 27.8 | 0.093 | 0.035 | 23.3 | 1.49 | 0.09 |
| | | | 25.1 | 1.50 | 0.11 |
| | | | 26.9 | 1.42 | 0.11 |
| | | | 28.6 | 1.34 | 0.12 |
| | | | 30.4 | 1.36 | 0.14 |
| | | | 32.2 | 1.14 | 0.11 |
| | | | 34.0 | 1.10 | 0.12 |
| | | | 36,7 | 1.24 | 0.12 |
| | | | 42.1 | 1.31 | 0.12 |

**Abell 2124**

| r(″) | $Mg_2$ | $\Delta Mg_2$ | r(″) | $D_{4000}$ | $\Delta D_{4000}$ |
|---|---|---|---|---|---|
| | | | -59.1 | 1.38 | 0.18 |
| | | | -48.3 | 1.36 | 0.18 |
| | | | -43.9 | 1.50 | 0.22 |
| | | | -40.3 | 1.34 | 0.15 |
| | | | -37.6 | 1.29 | 0.20 |
| | | | -35.8 | 1.31 | 0.16 |
| | | | -34.0 | 1.38 | 0.16 |
| | | | -32.2 | 1.47 | 0.18 |
| | | | -30.4 | 1.59 | 0.18 |
| | | | -28.6 | 1.70 | 0.19 |
| | | | -26.9 | 1.65 | 0.16 |
| | | | -25.1 | 1.57 | 0.14 |
| | | | -23.3 | 1.74 | 0.15 |
| | | | -21.5 | 1.66 | 0.14 |
| -23.3 | 0.090 | 0.030 | -19.7 | 1.66 | 0.12 |
| -18.8 | 0.157 | 0.029 | -17.9 | 1.86 | 0.13 |
| -16.1 | 0.161 | 0.034 | -16.1 | 1.78 | 0.12 |
| -14.3 | 0.177 | 0.028 | -14.3 | 1.84 | 0.11 |
| -12.5 | 0.198 | 0.027 | -12.5 | 1.94 | 0.10 |
| -10.7 | 0.190 | 0.022 | -10.7 | 1.89 | 0.08 |
| -9.0 | 0.188 | 0.018 | -9.0 | 1.89 | 0.07 |
| -7.2 | 0.233 | 0.017 | -7.2 | 1.96 | 0.07 |
| -5.4 | 0.239 | 0.013 | -5.4 | 1.98 | 0.05 |
| -3.6 | 0.245 | 0.011 | -3.6 | 2.01 | 0.05 |
| -1.8 | 0.268 | 0.010 | -1.8 | 2.11 | 0.04 |
| 0.4 | 0.282 | 0.009 | 0.4 | 2.14 | 0.04 |
| 1.8 | 0.277 | 0.010 | 1.8 | 2.11 | 0.04 |
| 3.6 | 0.258 | 0.011 | 3.6 | 2.05 | 0.05 |
| 5.4 | 0.241 | 0.014 | 5.4 | 1.96 | 0.06 |
| 7.2 | 0.232 | 0.017 | 7.2 | 1.94 | 0.07 |
| 9.0 | 0.214 | 0.020 | 9.0 | 1.89 | 0.07 |
| 10.7 | 0.212 | 0.023 | 10.7 | 1.79 | 0.08 |
| 12.5 | 0.159 | 0.027 | 12.5 | 1.84 | 0.10 |
| 14.3 | 0.197 | 0.032 | 14.3 | 1.98 | 0.13 |
| 17.0 | 0.190 | 0.027 | 16.1 | 1.91 | 0.15 |
| 21.5 | 0.133 | 0.031 | 17.9 | 1.80 | 0.16 |
| | | | 19.7 | 1.83 | 0.19 |
| | | | 22.4 | 1.90 | 0.17 |
| | | | 26.9 | 1.67 | 0.14 |

**Abell 2199**

| r(″) | $Mg_2$ | $\Delta Mg_2$ | r(″) | $D_{4000}$ | $\Delta D_{4000}$ |
|---|---|---|---|---|---|
| | | | -68.0 | 1.51 | 0.29 |
| | | | -59.1 | 1.45 | 0.21 |
| | | | -53.7 | 1.58 | 0.20 |
| | | | -49.2 | 1.60 | 0.18 |
| | | | -45.6 | 1.59 | 0.18 |
| | | | -42.1 | 1.64 | 0.16 |
| | | | -38.5 | 1.55 | 0.13 |
| | | | -35.8 | 1.56 | 0.15 |
| | | | -34.0 | 1.51 | 0.12 |
| | | | -32.2 | 1.53 | 0.13 |
| | | | -30.4 | 1.58 | 0.13 |
| | | | -28.6 | 1.81 | 0.14 |
| | | | -26.9 | 1.68 | 0.12 |
| -35.8 | 0.149 | 0.031 | -25.1 | 1.76 | 0.11 |
| -26.9 | 0.163 | 0.026 | -23.3 | 1.76 | 0.10 |
| -22.4 | 0.195 | 0.025 | -21.5 | 1.75 | 0.10 |
| -19.7 | 0.210 | 0.029 | -19.7 | 1.85 | 0.10 |
| -17.9 | 0.227 | 0.024 | -17.9 | 1.85 | 0.09 |
| -16.1 | 0.248 | 0.021 | -16.1 | 1.76 | 0.08 |
| -14.3 | 0.262 | 0.019 | -14.3 | 1.83 | 0.08 |
| -12.5 | 0.263 | 0.016 | -12.5 | 1.86 | 0.08 |
| -10.7 | 0.267 | 0.014 | -10.7 | 1.96 | 0.07 |
| -9.0 | 0.283 | 0.012 | -9.0 | 1.99 | 0.07 |
| -7.2 | 0.287 | 0.010 | -7.2 | 1.98 | 0.06 |
| -5.4 | 0.293 | 0.009 | -5.4 | 2.00 | 0.06 |
| -3.6 | 0.284 | 0.008 | -3.6 | 2.07 | 0.06 |
| -1.8 | 0.300 | 0.008 | -1.8 | 2.07 | 0.06 |
| 0.4 | 0.297 | 0.007 | 0.4 | 2.09 | 0.06 |
| 1.8 | 0.303 | 0.008 | 1.8 | 2.10 | 0.06 |
| 3.6 | 0.297 | 0.008 | 3.6 | 2.02 | 0.06 |
| 5.4 | 0.279 | 0.009 | 5.4 | 1.98 | 0.06 |
| 7.2 | 0.269 | 0.010 | 7.2 | 1.99 | 0.06 |
| 9.0 | 0.248 | 0.010 | 9.0 | 1.95 | 0.07 |
| 10.7 | 0.240 | 0.012 | 10.7 | 1.88 | 0.07 |
| 12.5 | 0.231 | 0.013 | 12.5 | 1.92 | 0.07 |
| 14.3 | 0.239 | 0.014 | 14.3 | 1.92 | 0.08 |
| 16.1 | 0.230 | 0.018 | 16.1 | 1.84 | 0.08 |
| 17.9 | 0.229 | 0.022 | 17.9 | 1.86 | 0.09 |
| 19.7 | 0.210 | 0.027 | 19.7 | 1.83 | 0.10 |
| 22.4 | 0.201 | 0.024 | 21.5 | 1.68 | 0.10 |
| 26.9 | 0.219 | 0.029 | 23.3 | 1.65 | 0.11 |
| 35.8 | 0.215 | 0.036 | 25.1 | 1.77 | 0.13 |
| | | | 26.9 | 1.56 | 0.12 |
| | | | 28.6 | 1.73 | 0.16 |
| | | | 30.4 | 1.59 | 0.15 |
| | | | 32.2 | 1.69 | 0.19 |
| | | | 34.0 | 1.64 | 0.20 |
| | | | 35.8 | 1.47 | 0.18 |
| | | | 38.5 | 1.56 | 0.17 |
| | | | 42.1 | 1.65 | 0.22 |
| | | | 49.2 | 1.58 | 0.19 |

**Abell 2319**

| r(″) | $Mg_2$ | $\Delta Mg_2$ | r(″) | $D_{4000}$ | $\Delta D_{4000}$ |
|---|---|---|---|---|---|
| | | | -42.1 | 1.35 | 0.14 |
| | | | -33.1 | 1.38 | 0.13 |
| | | | -29.5 | 1.43 | 0.12 |

**Table 5.** – *continued*

**Abell 2319 (contd.)**

| r(″) | $Mg_2$ | $\Delta Mg_2$ | r(″) | $D_{4000}$ | $\Delta D_{4000}$ |
|---|---|---|---|---|---|
| | | | -26.9 | 1.35 | 0.12 |
| | | | -25.1 | 1.48 | 0.13 |
| | | | -23.3 | 1.68 | 0.17 |
| | | | -21.5 | 1.80 | 0.14 |
| | | | -19.7 | 1.72 | 0.11 |
| -21.5 | 0.215 | 0.028 | -17.9 | 1.76 | 0.10 |
| -17.0 | 0.224 | 0.025 | -16.1 | 1.70 | 0.09 |
| -14.3 | 0.269 | 0.030 | -14.3 | 1.68 | 0.09 |
| -12.5 | 0.265 | 0.025 | -12.5 | 1.85 | 0.08 |
| -10.7 | 0.230 | 0.021 | -10.7 | 1.83 | 0.07 |
| -9.0 | 0.218 | 0.018 | -9.0 | 1.84 | 0.07 |
| -7.2 | 0.239 | 0.017 | -7.2 | 1.96 | 0.07 |
| -5.4 | 0.242 | 0.014 | -5.4 | 1.95 | 0.06 |
| -3.6 | 0.266 | 0.012 | -3.6 | 2.01 | 0.05 |
| -1.8 | 0.281 | 0.011 | -1.8 | 2.04 | 0.05 |
| 0.4 | 0.297 | 0.010 | 0.4 | 2.02 | 0.04 |
| 1.8 | 0.303 | 0.011 | 1.8 | 2.01 | 0.05 |
| 3.6 | 0.295 | 0.013 | 3.6 | 2.05 | 0.06 |
| 5.4 | 0.285 | 0.013 | 5.4 | 1.98 | 0.06 |
| 7.2 | 0.280 | 0.014 | 7.2 | 1.98 | 0.06 |
| 9.0 | 0.284 | 0.018 | 9.0 | 1.89 | 0.07 |
| 10.7 | 0.296 | 0.025 | 10.7 | 1.78 | 0.08 |
| 12.5 | 0.284 | 0.029 | 12.5 | 1.72 | 0.09 |
| 15.2 | 0.241 | 0.026 | 14.3 | 1.75 | 0.11 |
| 19.7 | 0.182 | 0.027 | 16.1 | 1.78 | 0.12 |
| | | | 17.9 | 1.70 | 0.12 |
| | | | 19.7 | 1.81 | 0.16 |
| | | | 21.5 | 1.67 | 0.14 |
| | | | 31.3 | 1.34 | 0.13 |

**Abell 2634**

| r(″) | $Mg_2$ | $\Delta Mg_2$ | r(″) | $D_{4000}$ | $\Delta D_{4000}$ |
|---|---|---|---|---|---|
| | | | -20.6 | 2.11 | 0.19 |
| | | | -17.4[b] | 1.97 | 0.15 |
| -16.8[b] | 0.214 | 0.030 | -16.1[b] | 2.24 | 0.14 |
| -15.1[b] | 0.271 | 0.027 | -15.1[b] | 2.17 | 0.10 |
| -14.1[b] | 0.292 | 0.019 | -14.1[b] | 2.05 | 0.05 |
| -13.1[b] | 0.261 | 0.014 | -13.1[b] | 2.14 | 0.04 |
| -12.1[b] | 0.278 | 0.014 | -12.1[b] | 2.10 | 0.04 |
| -11.1[b] | 0.278 | 0.015 | -11.1[b] | 2.12 | 0.04 |
| -10.1[b] | 0.260 | 0.019 | -10.1[b] | 2.08 | 0.06 |
| -9.1[b] | 0.254 | 0.025 | -9.1[b] | 2.02 | 0.08 |
| -8.0[b] | 0.201 | 0.031 | -8.0[b] | 2.14 | 0.10 |
| -7.0 | 0.272 | 0.032 | -7.0 | 1.98 | 0.11 |
| -6.0 | 0.239 | 0.030 | -6.0 | 2.10 | 0.11 |
| -5.0 | 0.239 | 0.026 | -5.0 | 2.14 | 0.09 |
| -4.0 | 0.264 | 0.022 | -4.0 | 2.05 | 0.07 |
| -3.0 | 0.272 | 0.017 | -3.0 | 2.12 | 0.06 |
| -2.0 | 0.307 | 0.014 | -2.0 | 2.11 | 0.04 |
| -1.0 | 0.289 | 0.012 | -1.0 | 2.20 | 0.03 |
| 0.3 | 0.291 | 0.011 | 0.3 | 2.20 | 0.03 |
| 1.0 | 0.276 | 0.012 | 1.0 | 2.24 | 0.04 |
| 2.0 | 0.289 | 0.014 | 2.0 | 2.17 | 0.04 |
| 3.0 | 0.272 | 0.018 | 3.0 | 2.19 | 0.06 |
| 4.0 | 0.242 | 0.023 | 4.0 | 2.17 | 0.08 |
| 5.0 | 0.247 | 0.030 | 5.0 | 2.13 | 0.11 |
| 6.4 | 0.292 | 0.030 | 6.0 | 1.87 | 0.10 |
| 9.4 | 0.273 | 0.031 | 7.0 | 2.12 | 0.17 |
| | | | 8.4 | 1.80 | 0.12 |
| | | | 10.7 | 1.88 | 0.14 |
| | | | 15.9 | 2.10 | 0.17 |

Note: $\Delta Mg_2$ and $\Delta D_{4000}$ are the formal errors in $Mg_2$ and $D_{4000}$ respectively. When necessary, the superscript "b" indicates the radii where the brightness of secondary nuclei dominates over the main galaxy profiles. For A 1126, the superscript "c" denotes a companion galaxy. The symbol † has been employed to denote the measurements obtained in Run 1 with the IPCS II detector.

**Table 6.** Line-strength gradients.

| Cluster | $G_{Mg_2}$ | $G_{D_{4000}}$ | $G^{int}_{Mg_2}$ | $G^{int}_{D_{4000}}$ |
|---|---|---|---|---|
| A 262 | −0.036 | −0.175 | −0.036 | −0.175 |
| | 0.015 | 0.052 | 0.015 | 0.052 |
| A 1126 | −0.035 | −0.086 | −0.026 | −0.012 |
| | 0.018 | 0.109 | 0.024 | 0.133 |
| A 1795 | −0.052 | −0.115 | −0.037 | −0.004 |
| | 0.012 | 0.045 | 0.009 | 0.031 |
| A 2124 | −0.109 | −0.407 | −0.096 | −0.290 |
| | 0.014 | 0.050 | 0.013 | 0.032 |
| A 2199 | −0.078 | −0.380 | −0.063 | −0.255 |
| | 0.012 | 0.032 | 0.011 | 0.031 |
| A 2319 | −0.058 | −0.417 | −0.056 | −0.311 |
| | 0.020 | 0.056 | 0.021 | 0.044 |
| A 2634 | −0.073 | −0.219 | −0.106 | −0.175 |
| | 0.032 | 0.131 | 0.036 | 0.170 |

Note: formal errors are given below data.

mean). A similar mean gradient of $-0.29 \pm 0.08$ is obtained for Davidge & Clark's sample, which comprises 5 elliptical galaxies (we have excluded data for the peculiar galaxy NGC 4486). The mean gradient in elliptical galaxies is then comparable to our mean $D_{4000}$ total gradient but larger, in absolute value, than that found in the inner regions, especially for cooling-flow galaxies. In fact, the mean internal gradient is shallower than those of Munn (1992) and Davidge et al. (1994) with confidence levels of $\sim 97\%$ and $\sim 94\%$ respectively (two-sample $t$ statistic test).

In summary, $Mg_2$ and $D_{4000}$ gradients in the central regions of galaxies with cooling flows seem to be shallower than those found in normal elliptical galaxies. This result is shown graphically in Figure 8, where we plot inner and total gradients against mass accretion rate. In Figure 8a we have included PKS 2354−35, with $G^{int}_{Mg_2} = -0.021$ (from GEA) and $\dot{M} = 320$ (from Schwartz et al. 1991). The dotted lines represent the mean gradients found in elliptical galaxies, being these larger than the internal gradients for almost all the cooling flow galaxies. Furthermore, these internal gradients (Figure 8a–8b) show some trend with accretion rate, in the sense that gradients flatten as $\dot{M}$ increases. Using the Spearman rank correlation test (Siegel 1956), these correlations are significant at the 97% and 90% levels, for $G^{int}_{Mg_2}$ and $G^{int}_{D_{4000}}$ respectively. It must be noted, however, that galaxies with moderate or null accretion rates span a wide range in index gradients and more high accretion rate galaxies should be observed to confirm the correlation. In addition it should not be forgotten that mass accretion rates are uncertain by about a factor of 2 (Arnaud 1988).

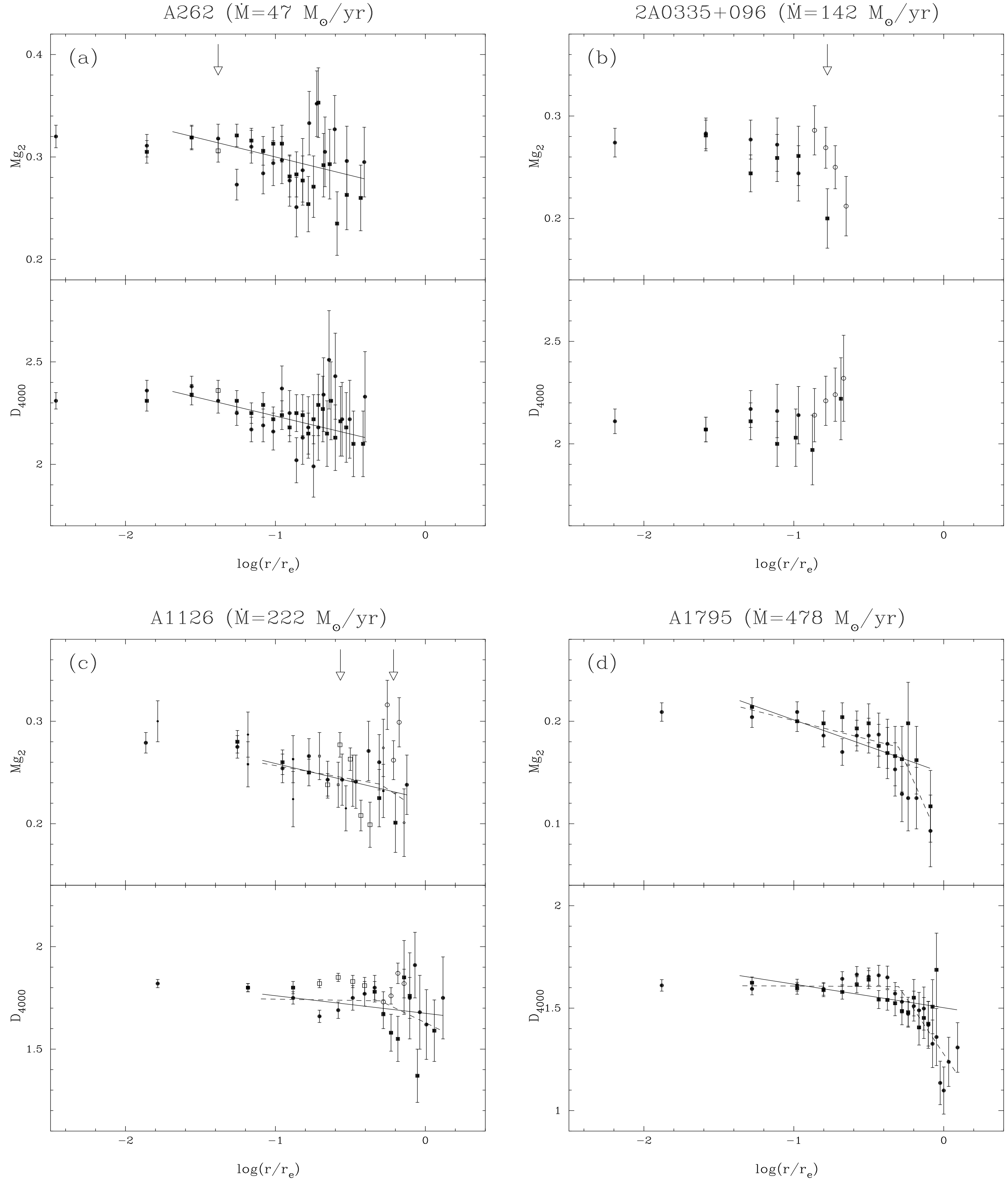

**Figure 7.** $Mg_2$ and $D_{4000}$ gradients for the galaxies of the sample. Filled circles and squares refer to different sides of the galaxies. When present, the position of the secondary nuclei is indicated by an arrow and open symbols have been employed at those radii where the brightness of the secondary nuclei dominates over the main galaxy profile. IPCS II $Mg_2$ data for the dominant galaxy in A 1126 are shown with small filled symbols. Full lines represent error-weighted least-squares fits to all the points, excluding either data with $r < 1.5''$ (affected by seeing) and secundary nuclei. Internal ($1.5'' < r \leq r_e/2$) and external ($r > r_e/2$) gradients were also fitted (dashed lines) and forced to join at $r = r_e/2$. Due to the low S/N ratio in the outer parts of the dominant galaxy in 2A 0335+096, fits were not computed for this object.

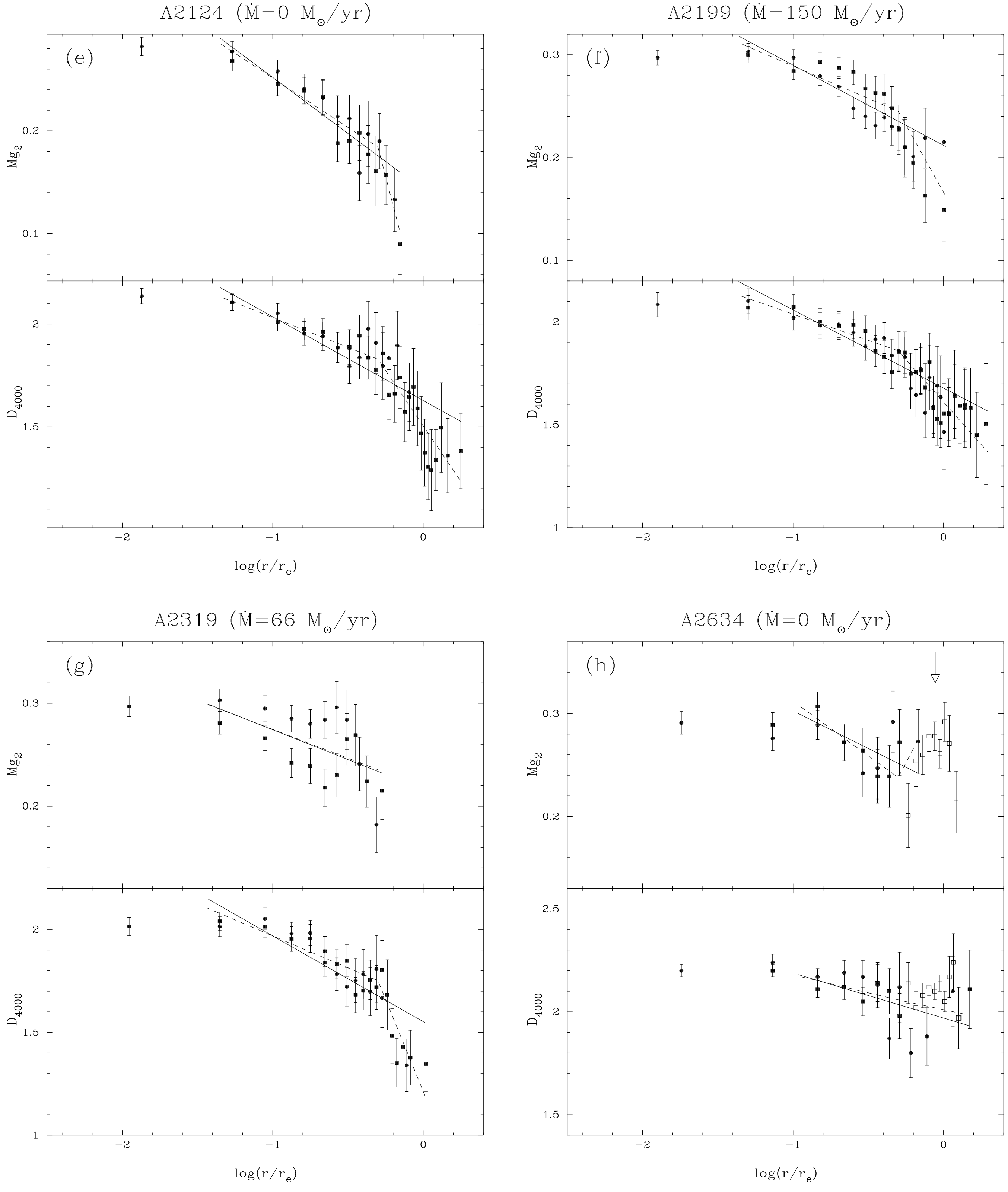

**Figure 7.** – *continued*

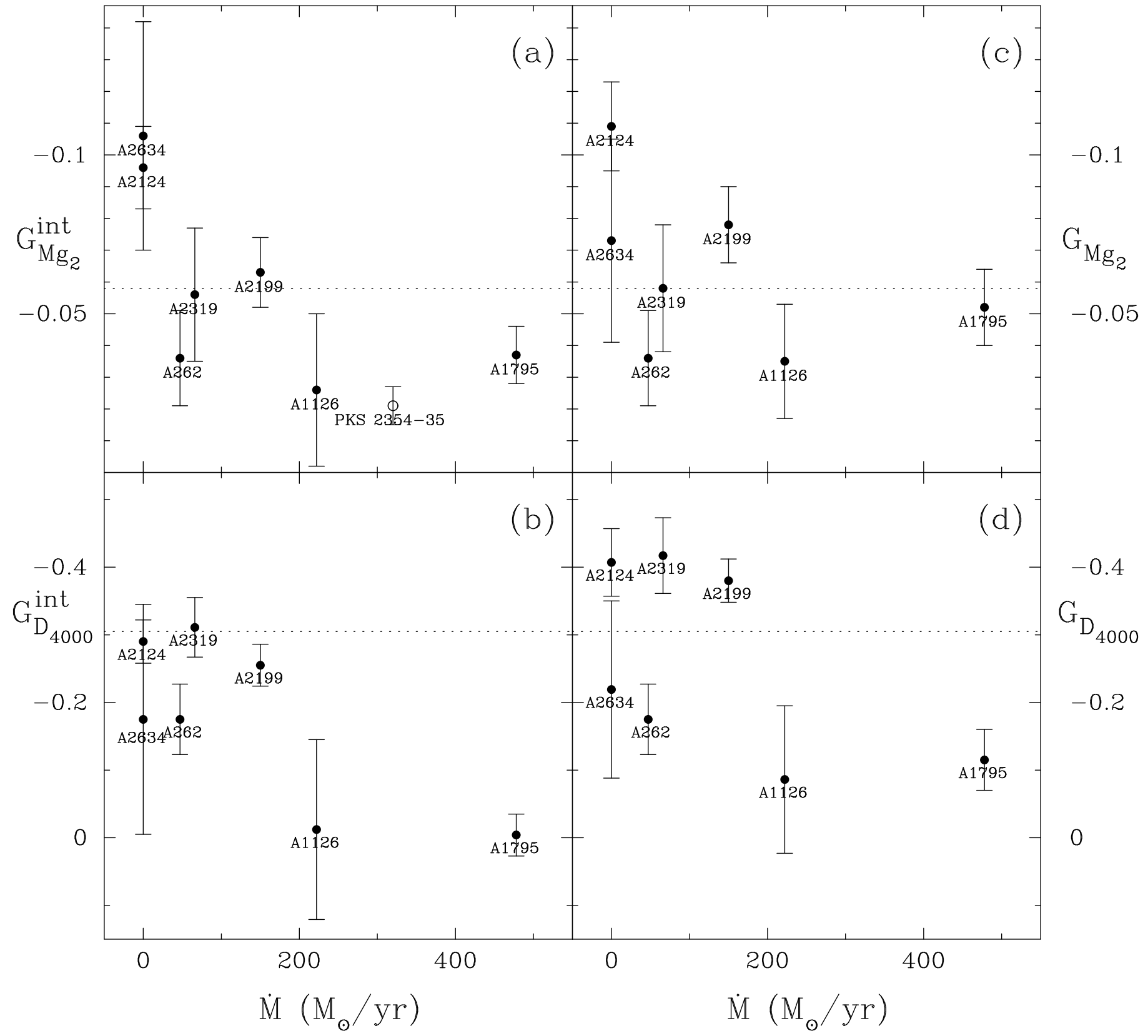

**Figure 8.** Measured internal ($G^{int}_{Mg_2}$, $G^{int}_{D_{4000}}$) and global ($G_{Mg_2}$, $G_{D_{4000}}$) gradients (Table 6) against mass accretion rate. In figure (a) we have included PKS 2354-35 from GEA. Dotted lines represent the mean spectral gradients in elliptical galaxies. The mean $Mg_2$ gradient is taking from GEA, whereas the $D_{4000}$ value is an average from the samples of Munn (1992) and Davidge & Clark (1994).

It is important to indicate that, if spectral gradients in the inner regions of cooling flow galaxies are really smaller than those of normal ellipticals, the aperture corrections applied in GEA are overestimated. This means that they did in fact find lower central $Mg_2$ indices in cooling flow objects. Besides, the central $D_{4000}$ indices of JFN would be hardly affected by aperture effects. The flat gradients found here confirms that the correlation of JFN is real, in agreement with our results of Section 3.3.

The interpretation of the measured $Mg_2$ and $D_{4000}$ gradients in the scenario of recent star formation induced by cooling flows is not straightforward. The models discussed in Section 3.2 (plotted in Figure 5), show that both $Mg_2$ and $D_{4000}$ decrease with the star formation rate. This implies that if stars were preferentially forming in the nuclear regions of the galaxies, the initial gradient of the underlying old stellar population (assuming it to be similar to that of normal ellipticals) would be smeared. This is in fact what we observed in our sample of cooling flow galaxies and, furthermore, the possible correlation of internal gradient with $\dot{M}$ suggests that star formation is concentrated in the inner regions.

However, in a less naive interpretation, the radial distribution of the expected changes in the spectral indices depends on the radial profile of the local ratio between the star formation rate and the luminosity of the underlying population. Assuming a drop-out parameter $n \sim 1$ (where $\dot{M}(<r) \propto r^n$) and that gas is immediately converted to stars at the dropout radius, we would expect the above ratio to increase roughly linearly with $r$ (O'Connell & McNamara 1989), and, therefore, the line-strength gradients in cooling flow galaxies would get strengthened, and they would increase with mass accretion rate, contrary to our findings.

Note however that there is no *a priori* reason to think that the star formation profile must coincide with the gas deposition profile. Obviously, the effectiveness of star formation rate, and even the shape of the IMF, may be radius dependent. Our results imply that, despite the fact that $\dot{M}$ can increase with $r$, the star formation rate, with a normal IMF, is much more efficient in the inner galaxy regions, probably due to the particular physical conditions of the cooled gas in the centres. Thus, since the amount of gas accreted by these central regions is a small fraction of the total gas flow, the derived star formation rate, when averaging over the whole galaxy, would only constitute a small fraction of the total X-ray derived $\dot{M}$. We do not find then a repository for the bulk of the accreted gas (i.e., the gas accreted at large radii compared with the optical size of the galaxy). The fate of this mass is still unknown, although some alternatives, like low-mass stars, have been proposed (Section 1.2 and references therein). Gorgas (1987) studied the expected variations of the $Mg_2$ index in a cooling flow galaxy as a function of the upper mass cutoff ($M_U$) of the IMF, finding, not surprisingly, that the star formation would be undetectable if $M_U \leq 0.6 M_\odot$.

Note that to obtain the SFR values of Section 3.3, we have implicitly used the hypothesis that $f_V$ (fraction of V luminosity that comes from the accretion population) is constant along the radius. In Figure 9 we investigate in more detail this assumption by plotting the expected variations in the mean $Mg_2$ and $D_{4000}$ gradients of a normal elliptical galaxy which is hosting a continuous star formation ($t = 5$ Gyr and Miller–Scalo IMF). For the underlying old population we have assumed the same central values used in Figure 5, whereas the mean gradients of the elliptical are taken from GEA and Munn (1992). Note that our models are still valid outside the galaxy centres since the intrinsic radial variations in metallicity (or even age) of the underlying population is taken into account by using the elliptical observed gradients. It is shown that gradients flatten as $f_V$ increases. Overimposed to these model predictions, we plot the observed gradient for A 1795, the galaxy with the highest accretion rate in our sample. When possible, indices have been averaged at both sides of the galaxy in order to obtain a smooth trend. The observations are consistent with a roughly constant $f_V$ (between 0.5 and 0.7 depending on the index considered) from the centre out to $r \approx r_e/2$.

Beyond that point $f_V$, as read from the graph, seems to grow dramatically and, in the outer regions, the $D_{4000}$ indices can not be explained as the result of a continuous SFR superimposed on an old (i.e. elliptical galaxy type) population. This apparent increase of $f_V$ is probably not real, and produced by other effects, since in the brightest galaxy of A 2124 ($\dot{M} = 0$) the indices in the outer parts attain quite similar values to those found for A 1795 (see below). Interestingly, the fact that $f_V$ seems to remain fairly constant in the inner regions of the galaxies suggests that the ongoing star formation does not follow the mass deposition profile inferred from X-ray data, in which case $f_V$ would grow roughly linearly with $r$, as argued above. The star formation is more concentrated towards the centre of the galaxy.

The low spectral indices measured in the outer regions of some galaxies (A 1795, A 2124, A 2319) deserve further attention. Whether this is a general property of the brightest cluster galaxies should be checked with further observations. Anyway, the fact that A 2124 is not a cooling-flow cluster suggests that a process not related to the cooling flows could be affecting the outer regions of central dominant galaxies. The interpretation of this result is not straightforward. The $Mg_2$ index by itself is not able to break the degeneracy between age and metallicity (Worthey 1994). If the $D_{4000}$ index were not sensitive to metallicity, the application of the Bruzual & Charlot (1993) models (restricted to solar metallicity) would imply an age of $\sim$ 1 Gyr for the bulk of the stellar population in these regions. On the other hand, it is still possible that these indices are due to an extremely low metallicity population, since Worthey's models suggest that $D_{4000}$ does have a strong metallicity dependency. Note, however, that Worthey's $D_{4000}$ calculations are based in model atmospheres (unlike for other spectral indices, where he uses observed data), and are, therefore, more uncertain. A firm conclusion can not be extracted until the dependence of $D_{4000}$ with the parameters of a composite population is investigated using observational stellar data. Anyway, it is important to note that normal elliptical galaxies do not usually attain these low indices at large radii (Two important exceptions are NGC 5813, Efstathiou & Gorgas 1985, and NGC 7144, Saglia et al. 1993, with $Mg_2$ gradients remarkably similar to A 2124). Although one could think of mechanisms that would lead to extremely low metallicities (when compared to ellipticals) in the outer parts of central dominant galaxies, our results suggest that a young stellar population may be present in the outer regions of brightest cluster galaxies. The above conclusion would not be surprising since central dominant galaxies could well be the sites of frequent merger/cannibalism processes and encounters with other cluster galaxies, which can easily trigger some kind of star formation phenomena. In this case, a mechanism able to produce star formation preferably in the outer galaxy regions, rather than in the centres, should be found.

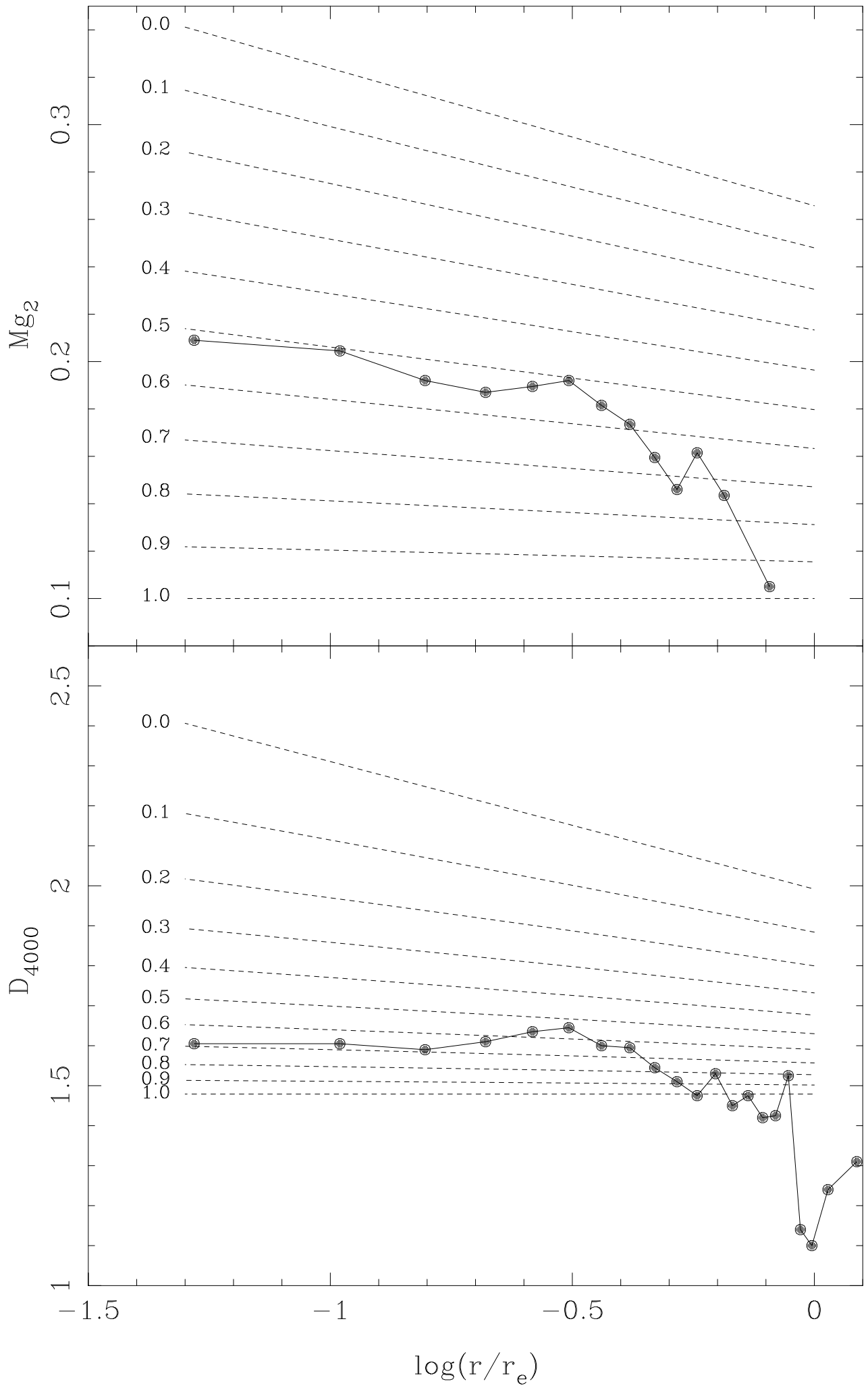

**Figure 9.** Dashed lines represent the expected variations of the mean $Mg_2$ and $D_{4000}$ gradients (from GEA and Munn 1992, respectively), as a function of $f_V$ (indicated by the numbers). Overimposed to these predictions, we have plotted the measured gradient for the dominant galaxy in A 1795 (where possible, indices have been averaged at both sides of the galaxy).

## 5 EMISSION LINES

Cooling flow galaxies usually exhibit prominent low-ionisation emission lines in their optical spectra and extended emission-line structures (Heckman 1981; Cowie et al. 1983; Hu et al. 1985; JFN; Heckman et al. 1989; Sparks, Macchetto & Golombek 1989; Shields & Filippenko 1990; Crawford & Fabian 1992; Allen et al. 1992). Employing line-ratio diagrams, Heckman et al. (1989) classified the measured emission into Types I and II, corresponding to different ionisation states of the gas, and found that this classification correlates with the X-ray and H$\alpha$ luminosities. In addition, Crawford & Fabian (1992) and Allen et al. (1992) showed that some objects exhibit both types of ionisation.

While some amounts ($\sim 10^5$–$10^8$ $M_\odot$) of warm gas (T$\sim 10^4$ K) have been detected in a large number of cooling flow cluster galaxies, many problems arise when trying to connect emission lines with cooling flows. The main difficulties, reviewed by Baum (1992) and Fabian (1994), can be summarised as follows: (1) optical emission lines are found in cooling flow galaxies, but they *are not* present in all such objects; (2) emission appears concentrated to the inner 10 kpc, although, if $\dot{M}(< r) \propto r$, most of the cooled gas should not be flowing into the cluster centre; (3) the ICM does not seem able to originate the observed optical emission filaments, with extents of $\sim$ 10 kpc; (4) H$\alpha$ luminosity is too high (by several orders of magnitude) to be explained as simple recombination of gas cooling at the mass accretion rate determined by X-ray measurements. Through the analysis of medium-resolution red spectra of the emission-line nebulae at the centres of four cooling flows, Donahue and Voit (1993) found that line-emitting filaments are not strong [Ca II] emission-line sources. These authors argued that this lack of [Ca II] emission indicates that the filaments may be dusty and, therefore, the nebular gas has *never* (their italics) been part of the hot ICM (dust immersed in the hot gas will be destroyed on a timescale of $\sim 10^6$ years). This implies that, though it seems clear that emission-line nebulae are linked to the existence of cooling flows, they would not be a direct product of them. However, Fabian, Johnstone & Daines (1994) note that dust may form in cooled gas clouds from cooling flows, so the argument of Donahue and Voit (1993) may not be correct.

The blue continuum detected in some massive cooling flow galaxies appears to be correlated with the Balmer emission line luminosities (JFN; Allen et al. 1992). However, photoionisation by young stars can not reproduce the measured line ratios, being necessary a much harder UV spectrum than that generated by hot stars. Voit & Donahue (1990) and Donahue & Voit (1991) proposed that soft X-ray radiation could be the ionisation mechanism but Crawford & Fabian (1992) could not reproduce, under this assumption, the transition from class I to class II nebulae. These last authors argued that Type I and Type II emission are produced by self-absorbed mixing layers and cloud-cloud collisions, respectively. Both mechanisms are related to the existence of cold clouds embedded in turbulent hot gas.

In this work we have not carried out a comprehensive study of the emission lines present in the galaxy spectra. We detect emission lines in 6 out of the 8 galaxies with cooling flows, being A 644 ($\dot{M} = 326 M_\odot$/yr) and A 2319 ($\dot{M} = 66 M_\odot$/yr) the objects without emission. These features are also seen in the spectra of the galaxy in A 2634, a non-cooling flow cluster. Note, however that this object is a strong radio source (3C 465) and its emission line system has already been studied by De Robertis & Yee (1990), concluding that the galaxy contains a weak AGN-like nucleus.

Since our spectra were not calibrated in absolute fluxes, we are not able to quantify the luminosity contained in the emission lines. However, we have measured the radial variation of the equivalent width of the H$\alpha$+[NII] system, and its flux (in arbitrary units). Both measures are plotted in Figure 10 for the galaxies with detected emission lines, exception made for PKS 0745−191 with too low S/N ratios. In addition, multi-component Gaussian profiles were fitted in the central spectra in order to obtain an estimate of the nuclear log([NII]$\lambda$6583/H$\alpha$) ratio. These values are given in the upper right corner of each plot in Figure 10 (except for A 2634, where the red spectral region does not have good enough S/N to obtain reliable Gaussian fits). Comparing our [NII]/H$\alpha$ ratios with those of Heckman et al. (1989), we obtain a very good agreement for the two galaxies in common. According to these ratios, we confirm that A 1795 (the cluster with the largest accretion rate) contains a Type II emission nebulae whereas the emission in the rest of the galaxies correspond to Type I, in the classification scheme of Heckman et al.

It is apparent from Figure 10 that the emission is concentrated in the inner galaxy regions, although in the case of 2A 0335+096 the emission is more extended. Simultaneously, some objects exhibit symmetric profiles (e.g. A 1126, A 1795), whereas other galaxies show different emission at both sides of the galaxy centre (e.g. 2A 0335+096, A 262). These asymmetries are not surprising since, in the case of 2A 0335+096, the H$\alpha$ images taken by Romanishin & Hintzen (1988) revealed a complex distribution of ionised hydrogen centred on the dominant cluster galaxy.

The main goal of our study of the emission lines is to determine the spatial coverage of the emission nebulae, especially in the case of A 1795. The data plotted in Figure 10 reveal that, for this object, emission is detected out to $\log(r/r_e) \sim 0.5$, roughly the same radius where the dominant galaxy in A 1795 exhibits a significant change of slope in the gradients of $Mg_2$ and $D_{4000}$ (Figure 7d). If the flattening of the gradients in the inner regions of this galaxy is due to the formation of new stars, we may question whether this star formation is related to the existence of emitting gas. A similar behaviour can be guessed for the dominant galaxy in A 2199, although, in this case the emission nebulae is not as extended as in A 1795. This trend cannot be confirmed for the rest of the cooling flow galaxies with emission, since the index gradients are not so well determined.

If the relationship between star formation and emission nebulae does actually exist, two questions must be asked. First, assuming that, as it has been noted by different authors (eg. Crawford & Fabian 1992), photoionisation by young stars can not be the fundamental mechanism for the excitation of the gas, what is the relative importance of that photoionisation in the line formation? The second question is why the bulk of the star formation is taken place in the same galaxy regions where emission lines are detected? It would be possible that, though a direct link between emission and star formation may not exist, physical conditions

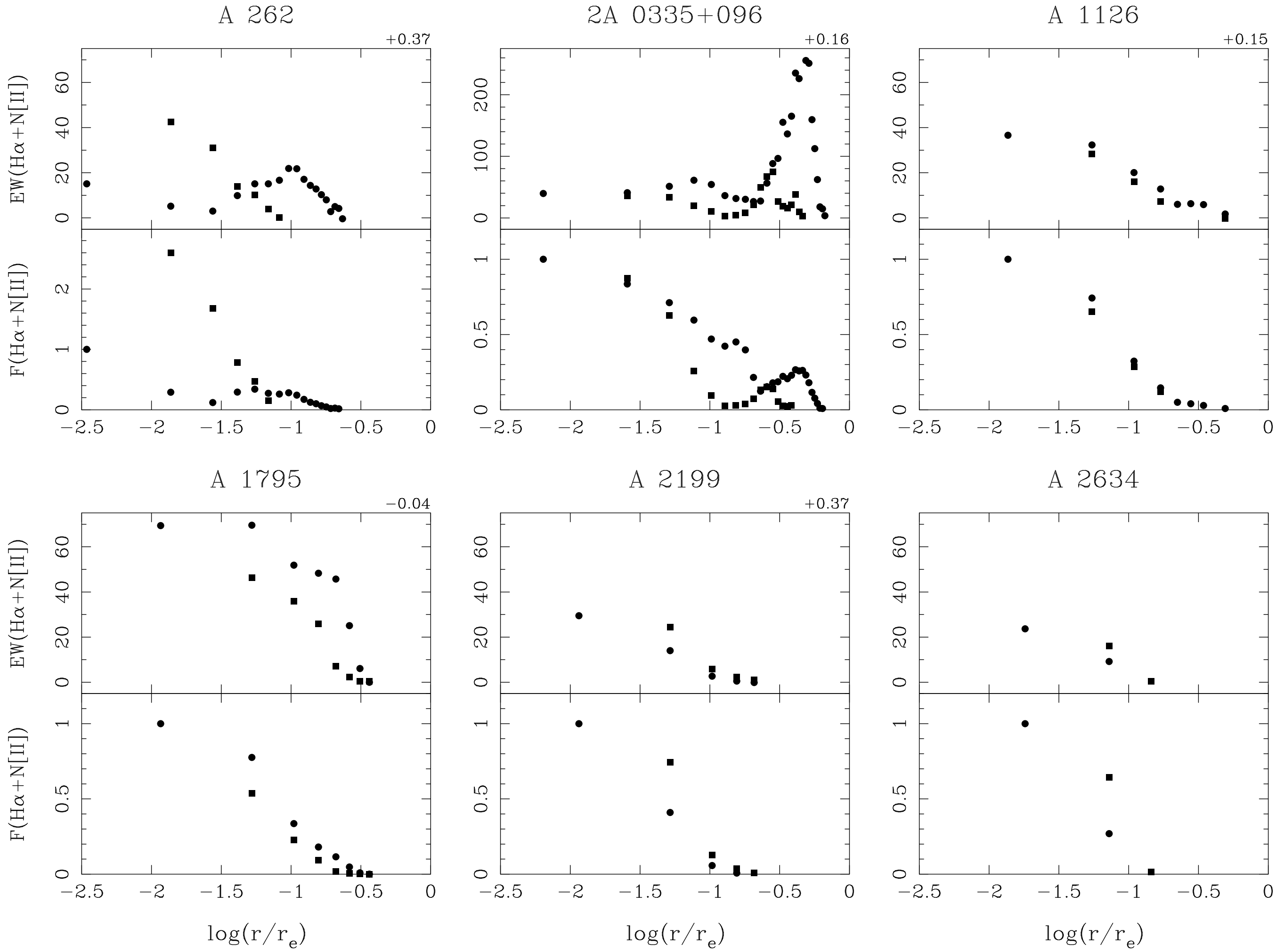

**Figure 10.** Equivalent width and flux (in relative units relative to the central emission) of the Hα+N[II] system as a function of galactocentric distance. Multi-component gaussian profiles were fitted in the central spectrum of each galaxy, and an estimate of log(N[II]λ6583/Hα) is given on the upper right corner of each plot (exception made for A 2634, with a central spectrum with too low S/N ratio in the Hα region). Squares and circles represent both sides of the galaxies (with the same criterion as in Figures 7).

in the cooling flows that produce one effect also favour the other.

We must also recall that the only object that clearly shows a Type II emission is the dominant galaxy of A 1795, which exhibits the strongest change in the slope of the gradients. In fact, the N[II]/Hα ratios seem to be correlated with the internal gradients, although we only have good data for four galaxies. Additional spectral gradients need to be measured in more galaxies in order to confirm this result. A relation of this type would imply that, since Type II emission is associated with large flows and therefore a larger central star formation could be expected, the nature of the emission lines, and its classification into different types, may also be related with young stars.

## 6 CONCLUSIONS

In this paper, the analysis of spectral indices in a sample of brightest cluster galaxies with and without cooling flows has led to the following conclusions:

i) We find that central $Mg_2$ and $D_{4000}$ indices are well correlated with the mass accretion rates derived from X-ray data, in the sense that indices decrease with $\dot{M}$. This supports the idea that cooling flow galaxies are hosting ongoing star formation from the accreted gas.

ii) Through the application of stellar population models we derive the Star Formation Rate implied by the observed spectral indices, which is also strongly correlated with the mass accretion rate. The fraction of the *total* accreted gas that is transformed into stars with a normal IMF is $5-17\%$ (depending on the duration of the star-formation phase) and almost the same for all galaxies. This small fraction is in agreement with previous studies. However, this is only a *lower limit* because only stars with $M > 0.1 M_{\odot}$ are considered. Moreover, since most of the mass deposition occurs at very large radii compared with the optical size of the galaxy, a large fraction (perhaps all) of the gas accreted *inside* the galaxy could be forming stars. We argue that this could provide a natural explanation for the very tight correlation between the derived SFR and $\dot{M}$ or, in other words, the almost constant SFR/$\dot{M}$ ratio. This fact, together with the lack of evidence for star formation at larger radii could be taken as an indication that only the gas that is being accreted *inside* the galaxy is converted into stars, perhaps due to the existence of very different physical conditions from those outside it§. We note here that our spectral indices do not place strong constraints to the shape of the IMF. Our measurements are consistent with a "normal" IMF and our estimated SFRs are valid for that case.

iii) For some cooling flow galaxies we find a clear slope change in the spectral gradients at intermediate radii. The derived gradients in the inner regions ($r \gtrsim r_e$) of cooling flow galaxies are, in the mean, shallower than those found in normal ellipticals. Furthermore, the data suggest a possible correlation of spectral gradients with mass accretion rate.

iv) The fraction of the V light that comes from the accretion population ($f_V$) is found to remain roughly constant with radius except, perhaps, for the outer parts of the galaxies. This suggests that the ongoing star formation does not follow the mass deposition profile inferred from X-ray data, being more concentrated towards the galaxy centre. This conclusion, however, depends on the assumption of a drop-out parameter $n \sim 1$, where $\dot{M}(< r) \propto r^n$ describes the gas deposition profile. Values of $n$ are generally poorly known and could vary considerably from cluster to cluster.

v) The indices reached in the outer regions of galaxies both with and without cooling flows are, in some cases, significantly lower than those usually found in normal ellipticals. We make the suggestion that recent star formation processes, not related with the cooling flows, could also be taking place in the outer parts of brightest cluster galaxies. We note however that further observations are needed to confirm this result.

vi) Emission lines, detected in 6 out of the 8 galaxies with cooling flows, appear to be concentrated in the inner galaxy regions. Furthermore, the emission region in A 1795 (and probably in A 2199) coincides approximately with the region where the flattening of spectral gradient is observed. If the flattening of the spectral gradients is related to the star formation induced by the cooling flows, perhaps this is not a coincidence. Evidently, a larger sample is needed.

Although we have shown that cooling flow galaxies exhibit spectral anomalies when compared with other brightest cluster galaxies, and that these anomalies correlate well with the size of the mass accretion rate. some words of caution are necessary. First, the sample should be enlarged to confirm some of the trends observed in this work, especially the low spectral indices measured in the outer regions of some galaxies. It should also be noticed that a cooling flow cD galaxy could be hosting star formation from the accreted cooled gas and, at the same time, be the site of additional star formation processes due to the cluster environment and evolution. Both processes would give rise to star formation that is both a function of time and galaxy radius, and thus the interpretation of spectral gradients in these galaxies is far from being straightforward.

§ In fact, Daines, Fabian & Thomas (1994), and Fabian (1994) note that cold clouds will weight the inner parts of a flow to the extent that they dominate the dynamics. In these regions (inside ~10 kpc) coagulation of clouds, shocks, mixing and massive star formation are possible (Fabian, private communication)

## ACKNOWLEDGEMENTS

We are indebted to A. C. Fabian for his comments and suggestions during the preparation of this paper. The WHT is operated on the island of La Palma by the Royal Greenwich Observatory at the Observatorio del Roque de los Muchachos of the Instituto de Astrofísica de Canarias. The Calar Alto Observatory is operated jointly by the Max-Planck-Institute für Astronomie (Heidelberg) and the Spanish Comisión Nacional de Astronomía. This work was supported in part by the Spanish "Programa Sectorial de Promoción General del Conocimiento" under grant No. PB93-456. AAS acknowledges financial support from the UK Particle Physics and Astronomy Research Council (PPARC), and the Royal Society.

## REFERENCES

Allen S. W., Edge A. C., Fabian A. C., Böhringer H., Crawford C. S., Ebeling H., Johnstone R. M., Naylor T., Schwarz R. A., 1992, MNRAS, 259, 67
Arnaud K. A., 1988, in Fabian A. C., ed, Proc. NATO Advanced Research Workshop, Cooling Flows in Clusters and Galaxies. Kluwer, Dordrecht, p. 31
Ball R., Burns J. O., Loken C., 1993, AJ, 105, 53
Baum S. A., 1992, PASP, 104, 848
Bertola F., Gregg M. D., Gunn J. E., Oemler A., 1986, ApJ, 303, 624
Bruzual A. G., 1983, ApJ, 273, 105
Bruzual A. G., Charlot S., 1993, ApJ, 405, 538
Burstein D., Heiles C., 1982, AJ, 87, 1165
Carollo C. M., Danziger I. J., 1994, MNRAS, in press
Carollo C. M., Danziger I. J., Buson L., 1993, MNRAS, 265, 553
Charlot S., Silk J., 1994, ApJ, in press
Cole S., Aragón-Salamanca A., Frenk C. S., Navarro J. F., Zepf S. E., 1994, MNRAS, 271, 781
Cowie L. L., Hu E. M., Jenkins E. B., York D. G., 1983, ApJ, 272, 29
Crawford C. S., Fabian A. C., 1992, MNRAS, 259, 265
Crawford C. S., Fabian A. C., 1993, MNRAS, 265, 431
Crawford C. S., Arnaud K. A., Fabian A. C, Johnstone R. M., 1989, MNRAS, 236, 277
Daines S. J., Fabian A. C., Thomas P. A., 1994, MNRAS, 268, 1060
Davidge T. J., Clark C. C., 1994, AJ, 107, 946
Davies R. L., Burstein D., Dressler A., Faber S. M., Lynden-Bell D., Terlevich R., Wegner G., 1987, ApJS, 64, 581
Davies R. L., Sadler E. M., Peletier R. F., 1993, MNRAS, 262, 650
De Jong T., Norgaard-Nielsen H. U., Jorgensen H. E., Hansen L., 1990, A&A, 232, 317
De Robertis M. M., Yee H. K. C., 1990, AJ, 100, 84
Donahue M., Voit G. M., 1991, ApJ, 381, 361
Donahue M., Voit G. M., 1993, ApJ, 414, L17
Dressler A., 1988, in Faber S. M., ed, Nearly Normal Galaxies. New York, Springer-Verlag, p. 276
Dressler A., Shectman S. A., 1987, AJ, 94, 899
Edge A. C., Stewart G. C., Fabian A. C., 1992, MNRAS, 258, 177
Efstathiou G., Gorgas J., 1985, MNRAS, 215, 37P
Faber S. M., Friel E. D., Burstein D., Gaskell C. M., 1985, ApJS, 57, 711
Fabian A. C., 1994, ARA&A, 32, 277
Fabian A. C., Nulsen P. E. J., Canizares C. R., 1984, Nat, 310, 733
Fabian A. C., Nulsen P. E. J., Canizares C. R., 1991, A&AR, 2, 191

Fabian A. C., Canizares C. R., Böhringer H., 1994, ApJ, 425, 40
Fabian A. C., Johnstone, R. M., Daines, S. J., 1994, MNRAS, 271, 737
Ferland G., Fabian A. C., Johnstone R. M., 1994, MNRAS, 266, 399
González J. J., 1993, PhD Thesis, University of California, Santa Cruz
Gorgas J., 1987, PhD Thesis, Universidad Complutense, Madrid
Gorgas J., Efstathiou G., Aragón-Salamanca A., 1990, MNRAS, 245, 217 (GEA)
Gorgas J., Faber S. M., Burstein D., González J. J., Courteau S., Prosser C., 1993, ApJS, 86, 153
Hamilton D., 1985, ApJ, 297, 371
Heckman T. M., 1981, ApJ, 250, L59
Heckman T. M., Baum S. A., van Breugel W. J. M., McCarthy P., 1989, ApJ, 338, 48
Henriksen M. J., 1993, ApJ, 407, L13
Hu E. M., Cowie L. L., Wang Z., 1985, ApJS, 59, 447
Johnstone R. M., Fabian A. C., 1989, MNRAS, 237, 27P
Johnstone R. M., Fabian A. C., Nulsen P. E. J., 1987, MNRAS, 224, 75 (JFN)
King D. L., 1985, in RGO/La Palma Technical Notes, No. 31
Lachièze-Rey M., Vigroux, L., Souviron, J., 1985, A&A, 150, 62
Maccagni D., Garilli B., Gioia I. M., Maccacaro T., Vettolani G., Wolter A., 1988, ApJ, 334, L1
Mackie G., Visvanathan N., Carter D., 1990, ApJS, 73, 637
Massey P., Strobel K., Barnes J. V., Anderson E., 1988, ApJ, 328, 315
McNamara B. R., O'Connell R. W., 1989, AJ, 98, 2018
McNamara B. R., O'Connell R. W., 1992, ApJ, 393, 579
McNamara B. R., O'Connell R. W., 1993, AJ, 105, 417
McNamara B. R., Bregman J. N., O'Connell R. W., 1990, ApJ, 360, 20
Miller G. E., Scalo J. M., 1979, ApJS, 41, 513
Munn J. A., 1992, ApJ, 399, 444
Mushotzky R. F., 1992, in Fabian A. C., ed, Clusters and Superclusters of Galaxies. Kluwer, Dordrecht, p. 91
O'Connell R. W., McNamara B. R., 1989, AJ, 98, 180
Oke J. B., 1974, ApJS, 27, 21
Peletier R. F., 1992, in Danziger I. J., Zeilinger W. W., Kjär K., eds, Proc. ESO/EIPC Workshop No. 45, Structure, Dynamics and Chemical Evolution of Elliptical Galaxies. ESO, Germany, p. 409
Romanishin W., 1986, ApJ, 301, 675
Romanishin W., 1987, ApJ, 323, L113
Romanishin W., Hintzen P., 1988, ApJ, 324, L17
Saglia R. P., Bertin G., Bertola F., Danziger J., Dejonghe H., Sadler E. M., Stiavelli M., de Zeeuw P. T., Zeilinger W. W., 1993, ApJ, 403, 567
Salpeter E. E., 1955, ApJ, 121, 161
Sarazin C. L., 1986, Rev. Mod. Phys., 58, 1
Sarazin C. L., O'Connell R. W., 1983, ApJ, 268, 552
Savage B. D., Mathis J. S., 1979, ARA&A, 17, 73
Scalo J. M., 1986, Fund. Cosmic Phys., 11, 1
Schombert J. M., 1986, ApJS, 60, 603
Schombert J. M., 1987, ApJS, 64, 643
Schombert J. M., Barsony M., Hanlon P. C., 1993, ApJ, 416, L61
Schwartz D. A., Bradt H. V., Remillard R. A., Tuohy, I. R., 1991, ApJ, 376, 424
Shields J. C., Filippenko A. V., 1990, ApJ, 353, L7
Siegel, S., 1956, Nonparametric Statistics. McGraw-Hill, Kogakusha.
Sparks W. B., 1992, ApJ, 399, 66
Sparks W. B., Macchetto F., Golombek D., 1989, ApJ, 345, 153
Thomas P. A., Fabian A. C., Nulsen P. E. J., 1987, MNRAS, 228, 973
Thuan T. X., Puschell J. J., 1989, ApJ, 346, 34
Voit G. M., Donahue M., 1990, ApJ, 360, L15
White D. A., Fabian A. C., Johnstone R. M., Mushotzky R. F., Arnaud K. A., 1991, MNRAS, 252, 72
Wirth A., Kenyon S. J., Hunter D. A., 1983, ApJ, 269, 102
Worthey G., 1994, ApJS, 95, 107